\documentclass[a4paper,fleqn,usenatbib]{mnras}
\usepackage{newtxtext,newtxmath}

\usepackage[T1]{fontenc}
\usepackage{ae,aecompl}
\usepackage{graphicx}	
\usepackage{amsmath}	
\usepackage{amssymb}	
\usepackage{hyperref}
\usepackage{microtype}
\usepackage{verbatim}
\usepackage{graphicx}
\usepackage{subfig}
\usepackage{url}

\title[]{\centering{A multi-wavelength view of distinct accretion regimes in the pulsating ultraluminous X-ray source NGC 1313 X-2}}

\author[R. Sathyaprakash et al.]{R. Sathyaprakash,$^{1,2}$\thanks{E-mail: sathyaprakash@ice.csic.es},
T.\,P. Roberts$^{1}$,
F.\ Gris\'e$^{3}$,
P.\ Kaaret$^{4}$,
E.\ Ambrosi$^{5}$,
C.\ Done$^{1}$,
\newauthor
J.\,C.\ Gladstone$^{6}$,
J.\,J.\,E. Kajava$^{7}$,
R.\ Soria$^{8,9}$,
L.\ Zampieri$^{10}$
\\
$^{1}$Centre for Extragalactic Astronomy \& Department of Physics, Durham University, South Road, Durham DH1 3LE, UK\\
$^{2}$Institute of Space Sciences (IEEC-CSIC), Campus UAB, Carrer de Can Margans, s/n 08193 Barcelona, Spain \\
$^{3}$The Pennsylvania State University, 525 Davey Lab, University Park, PA 16802, USA\\
$^{4}$Department of Physics and Astronomy, University of Iowa, Van Allen Hall, Iowa City, IA 52242, USA\\
$^{5}$INAF/IASF Palermo, via Ugo La Malfa 153, I-90146 - Palermo, Italy\\
$^{6}$ Department of Physics, University of Alberta, 11322-89 Avenue, Edmonton, AB T6G 2G7, Canada\\
$^{7}$Department of Physics and Astronomy, FI-20014 University of Turku, Finland\\
$^{8}$College of Astronomy and Space Sciences, University of the Chinese Academy of Sciences, Beijing 100049, China\\
$^{9}$Sydney Institute for Astronomy, School of Physics A28, The University of Sydney, Sydney, NSW 2006, Australia\\
$^{10}$INAF - Astronomical Observatory of Padova, Vicolo dell’Osservatorio 5, 35122, Padova, Italy
}
\date{Accepted 2022 February 11. Received 2022 February 11; in original form 2021 October 14}

\pubyear{2022}

\begin{document}
\label{firstpage}
\pagerange{\pageref{firstpage}--\pageref{lastpage}}
\maketitle

\begin{abstract}
\begin{centering}
NGC 1313 X-2 is one of the few known pulsating ultraluminous X-ray sources (PULXs), and so is thought to contain a neutron star that accretes at highly super-Eddington rates.  However, the physics of this accretion remains to be determined.  Here we report the results of two simultaneous {\it XMM-Newton\/} and {\it HST\/} observations of this PULX taken to observe two distinct X-ray behaviours as defined from its {\it Swift\/} light curve.  We find that the X-ray spectrum of the PULX is best described by the hard ultraluminous (HUL) regime during the observation taken in the lower flux, lower variability amplitude behaviour; its spectrum changes to a broadened disc during the higher flux, higher variability amplitude epoch.  However, we see no accompanying changes in the optical/UV fluxes, with the only difference being a reduction in flux in the near-IR as the X-ray flux increased.  We attempt to fit irradiation models to explain the UV/optical/IR fluxes but they fail to provide meaningful constraints.  Instead, a physical model for the system leads us to conclude that the optical light is dominated by a companion O/B star, albeit with an IR excess that may be indicative of a jet.  We discuss how these results may be consistent with the precession of the inner regions of the accretion disc leading to changes in the observed X-ray properties, but not the optical, and whether we should expect to observe reprocessed emission from ULXs.
\end{centering}
\end{abstract}


\section{Introduction}
\label{section:intro}

Ultraluminous X-ray sources (ULXs) are point-like off nuclear sources that are observed to radiate in excess of the Eddington limit for a ten solar mass black hole assuming isotropic emission (see \citealt*{r1} for a comprehensive review). It is suspected that the majority of the currently observable population of ULXs are powered by super-critical accretion (from a companion star) onto either neutron stars (\citealt{r2}; \citealt{r3,r4}; \citealt{r5}; \citealt{r6}; \citealt{r7}) or stellar mass black holes (\citealt{r8}), sometimes reaching luminosities up to $\sim 500$ L$_{\text{Edd}}$. However, the nature of the accretor has been confirmed in only a handful of objects that are observed to pulsate or those showing cyclotron line features thought to form in the vicinity of a neutron star's magnetosphere (see e.g. \citealt{r1g}), whilst the identity of all other ULXs remains an open question. 


The spectral and timing variability patterns of several ULXs are well explained by a super-critical wind model that traces its origins to the work of \citet{r10}. Briefly, when the mass accretion rate through the disc exceeds the local Eddington rate, the disc is predicted to become geometrically thick due to radiation pressure. As a consequence, some fraction of the inflowing mass is thought to escape as a wind, while much of the rest is advected on to the compact object (\citealt{r11}, \citealt{r12}, \citealt{r13}). Therefore, it is natural to expect the presence of an optically thick outflow to influence the observational appearance of super-Eddington ULXs. In particular, differences in both the viewing angle of the system and the mass accretion rate can explain the plethora of ULX spectral shapes, ranging from those seen in supersoft to hard ultraluminous sources (\citealt*{r14,r101}).  The significant detection of highly blue-shifted absorption lines in some ULXs (\citealt{r15,r16} and \citealt{r17,r18}) provides strong evidence for these radiatively driven winds.  It is notable that such winds could be present even if the compact object is a high magnetic field neutron star \citep{r102}.

Dedicated X-ray monitoring campaigns on some ULXs have revealed that they show quasi-periodic variations in the X-ray flux on timescales larger than $\sim$ 50 days (e.g. \citealt{r19}; \citealt{r20}; \citealt{r103}). These variations seem to disappear over sufficiently long baselines and are hence difficult to interpret as the orbital period of the system (e.g. \citealt{r1b}). Recently, \citet{r21} have reported that at least four ULXs display a strong quasi-periodic modulation in the hard X-ray band, with little variability at energies below 1 keV. These ULXs also show a bimodal distribution on the hardness-intensity plane, with the brightest sources appearing to have a slightly harder spectrum. This is consistent with the proposition that long-term trends in the chaotic X-ray variability are driven by the precession of a super-critical accretion flow (\citealt{r22,r23}), rather than an exclusive consequence of variation in the mass accretion rate. 

One of the objects that \citet{r21} demonstrate to show this behaviour is the archetypal ULX, NGC 1313 X-2.  This is one of the best-studied ULXs in X-rays, with the examination of its X-ray spectra the subject of many papers (e.g. \citealt{r104}; \citealt{r105}; \citealt{r108}).  Its spectrum is similar to that of some other bright ULXs with hard ultraluminous (HUL) spectra at moderate luminosities, which evolve to appear dominated by a single disc-like component similar to the broadened disc (BD) regime at peak luminosities (cf. \citealt{r106}; \citealt{r22}). Interestingly, the spectrum at high luminosity appears very similar to the dominant spectra observed in pulsating ULXs (PULXs) that host a neutron star \citep{r107}.  Indeed, NGC 1313 X-2 has been revealed to be a PULX, with the discovery of faint pulsations in recent long {\it XMM-Newton\/} observations reported by \citet{r6}.  Its phenomenology is therefore of utmost interest as we begin to observationally disentangle the mystery of the emission mechanisms in such objects.

NGC 1313 X-2 is also one of the best studied ULXs at optical wavebands. \citet{r24} identified two potential optical counterparts to the X-ray source, although improved {\it HST\/} and {\it{Chandra}} astrometry have shown that only one of these is a plausible candidate. The true counterpart is known to be variable in the optical waveband, with short timescale ($\sim$ days) periodic modulations initially reported by \citet{r25} (and later questioned by \citealt{r114}). They interpreted this as arising due to X-ray heating of the donor's surface, which can produce periodic variations in the observed intensity as the X-ray heated (more luminous) face of the star moves in and out of the observer's line-of-sight over a complete orbital cycle (provided the system is not viewed too close to face on; \citealt{r26}). However, these periodic variations were absent in subsequent observations of X-2, perhaps having been masked out by a stronger contribution from the accretion disc, which is expected to vary stochastically (\citealt{r27}). Further spectroscopic monitoring of the ULX counterpart with Gemini-South revealed a broad and variable He II $\lambda 4686$ emission line. The variations in its central wavelength were found to be non-periodic, thus unlikely to be associated with the orbital motion of the donor (\citealt{r28}). Instead, the variability is stochastic, suggesting an origin in the reprocessing of X-rays in the outer accretion disc or a radiatively accelerated outflow (\citealt{r27}; \citealt{r29}). 

Previous works analysing the broadband spectral energy distributions (SEDs) of ULXs have been unable to distinguish unambiguously between an irradiated disc and donor star models (e.g. \citealt{r30}). However, quasi-simultaneous X-ray and optical monitoring of X-2 showed that its B-band magnitude displays larger amplitude variability during periods of high X-ray flux levels (\citealt{r27}), giving some credence to irradiation scenarios. Therefore, studying correlated changes between the X-ray and optical emission could be the key to breaking the degeneracy in the emission mechanisms associated with the optical counterpart.  They could also provide further insight into the physical process(es) driving the transition between the bimodal hardness-intensity regimes in X-2 and by extension other ULXs showing similar behaviour. 

In order to investigate this problem further, we obtained simultaneous {\it{XMM-Newton}} and {\it HST\/} observations of NGC 1313 X-2.   In this paper we report the results of these observations.  We describe the observations in Section 2 and present the X-ray and optical results involving the modelling of the X-ray-to-optical Spectral Energy Distributions (SEDs) in Section 3.  We discuss these results in Section 4 before concluding the paper in Section 5.  

\begin{figure*}
\includegraphics[scale=0.8]{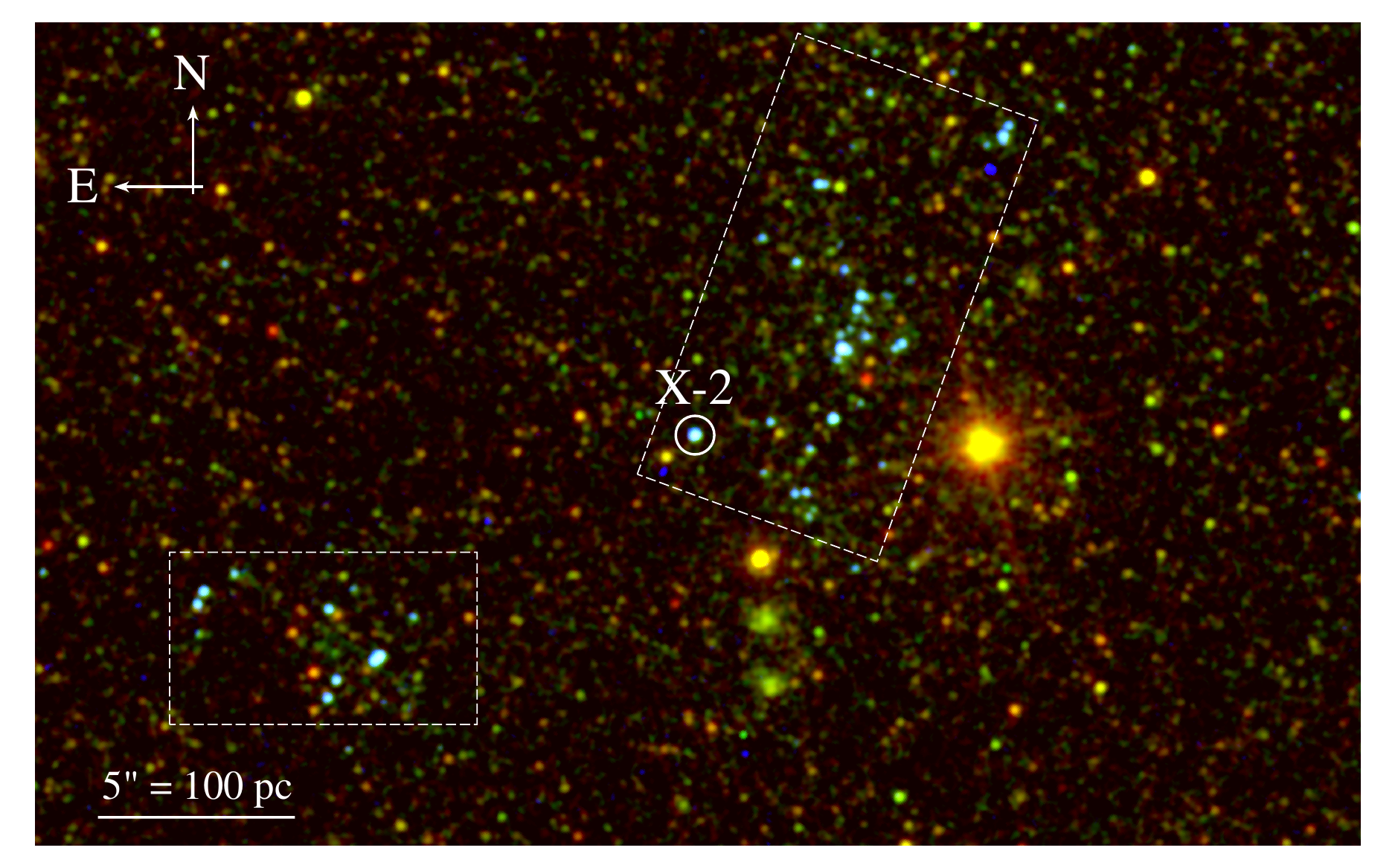}
\caption{An {\it HST}/WFC3 composite image of the field containing the ULX NGC 1313 X-2 (whose position is marked by a circle of radius 0.5 arcseconds). Red, green and blue colours represent intensities in the F814W, F555W and F225W filters, respectively. The dashed boxes highlight the approximate location of two OB associations in the vicinity of the ULX. The image has been smoothed with a Gaussian kernel with $\sigma = 1.5$ pixels ($\approx$ 0.06 arcseconds).}    
\label{image}
\end{figure*}

\section{Data reduction}
\label{section:data_red}

We obtained a multi-wavelength observation programme (PI: Gris\'e) with the aim of studying the near-IR through X-ray emission properties of NGC 1313 X-$2$ in two distinct phases of its behaviour (which will be defined in Section 3). This provided simultaneous X-ray spectroscopy of the ULX from the {\it XMM-Newton\/} EPIC cameras and photometry of the optical counterpart in several bands (ranging from the near-IR to near-UV) from the {\it HST\/} WFC3 and ACS cameras. We also use {\it{Swift}} data to study the long-term variability properties of the source in order to track its change in accretion state, which was used as a trigger to begin the {\it{XMM-Newton}} and {\it{HST}} observations. 


\subsection{XMM-Newton X-ray observations}
\label{subsection:xmm}

The \textit{XMM-Newton} observations were carried out with the EPIC-pn and MOS detectors in full-frame imaging mode, and were reduced using the Science Analysis Software (SAS)v.16.\footnote{https://www.cosmos.esa.int/web/xmm-newton/xsa} We summarise the essential features of these observations in Table~\ref{Tab:Xlog}. After examining the radial profile to confirm there were no extended components to the point spread function, we extracted source events within an aperture of 30 arcseconds centred on the \textit{Chandra} source position (RA = 49.592, DEC = -66.601; \citealt{r31}). The background region was chosen to lie adjacent to the source on the same chip. Care was taken to avoid areas with a large internal quiescent background in the EPIC-pn detector, which is primarily traced by the intensity of the Copper-K$\alpha$ emission line (at $\sim$ 8 keV). We excluded events associated with bad pixels by implementing the standard flagging criteria ({\tt{FLAG==0}}) and grade selection ({\tt{PATTERN}} $\leq 12$ for MOS and {\tt{PATTERN}} $\leq 4$ for PN). We extracted light curves in the 10 - 12 keV energy band using the {\tt{evselect}} task from chips devoid of a large number of point sources, in order to inspect and remove periods of intense soft proton flares. We ran the SAS task {\tt{bkgoptrate}} to determine a safe count rate threshold below which the background level remains steady. As such, periods during which the count rates exceeded this threshold (in the 10 - 12 keV band) were excluded; this resulted in a loss of up to $\sim 46\%$ of the original exposure for the pn camera in the March 24 observation, with other losses being smaller.  Finally, using the cleaned event files we extracted the source and background spectra in the 0.3 - 10.0 keV energy range for each instrument. The spectra were grouped into a minimum of 25 counts after background subtraction per energy bin, ensuring not to oversample the intrinsic detector resolution by a factor larger than 3. Hence, it was acceptable to use $\chi^{2}$ statistics to characterise the goodness-of-fit. Finally, in order to deconvolve the energy spectrum through the detector response, we generated the auxillary response files and redistribution matrix files using the {\tt{arfgen}} and {\tt{rmfgen}} tasks respectively. 

\subsection{HST optical, UV and IR observations}
\label{subsection:HST}

\begin{table*}
\caption{A log of the {\it{XMM-Newton}} X-ray observations.}
\centering
\begin{tabular}{c c c c}
\hline
Dataset & Start date & GTI-corrected exposure & Combined EPIC count-rate \\[0.5ex]
\newline 
\newline
(Obs. ID) & (dd-mm-yy) & (pn/MOS1/MOS2, in ks) & (s$^{-1}$) \\
\hline
0764770101 & 05-12-15 & 50.1/68.7/68.4 & 0.247 $\pm$ 0.002  \\
0764770401 & 24-03-16 & 16.4/21.5/21.0  & 0.422 $\pm$ 0.005 \\
\hline
\end{tabular}
\label{Tab:Xlog}
\end{table*}

\begin{table*}
\caption{{\it HST\/} optical, UV and IR magnitudes of the ULX counterpart.}
\centering
\begin{tabular}{c c c c c c c}
\hline
\hline
Start date & Instrument & Filter$^{\text{a}}$ & Exposure time & Aperture corrected magnitude$^{\text{b}}$$^{\text{c}}$ & Aperture correction$^{\text{c}}$\\[0.5ex]
(dd-mm-yy) & & & (s) &  (Vega mag) & (Vega mag) \\
\hline
05-12-15 & ACS/SBC & F140LP & 1200.0 & 20.56 $\pm$ 0.07 & 0.2\\
       & WFC/UVIS & F225W & 1440.0 & 21.34 $\pm$ 0.05 & 0.25 $\pm$ 0.04\\
       & WFC/UVIS & F336W & 1017.0 & 21.93 $\pm$ 0.08 & 0.24 $\pm$ 0.07\\
       & WFC/UVIS & F438W & 1050.0 & 23.35 $\pm$ 0.06 & 0.26 $\pm$ 0.05\\
       & WFC/UVIS & F555W & 1170.0 & 23.41 $\pm$ 0.05 & 0.69 $\pm$ 0.05\\
       & WFC/UVIS & F814W & 1740.0 & 23.46 $\pm$ 0.06 & 0.34 $\pm$ 0.06\\
       & WFC/IR & F125W & 1796.9 & 23.30 $\pm$ 0.08 & 0.41 $\pm$ 0.08\\
\hline
24-03-16 & ACS/SBC & F140LP & 1200.0 & 20.65 $\pm$ 0.07 & 0.2\\
	& WFC/UVIS & F225W & 1440.0 & 21.25 $\pm$ 0.05 & 0.18 $\pm$ 0.04\\
	& WFC/UVIS & F336W & 1017.0 & 21.90 $\pm$ 0.05 & 0.18 $\pm$ 0.04\\
	& WFC/UVIS & F438W & 1050.0 & 23.43 $\pm$ 0.05 & 0.19 $\pm$ 0.05\\
	& WFC/UVIS & F555W & 1170.0 & 23.50 $\pm$ 0.04 & 0.20 $\pm$ 0.04\\
	& WFC/UVIS & F814W & 1740.0 & 23.57 $\pm$ 0.05 & 0.29 $\pm$ 0.05\\
	& WFC/IR & F125W & 1796.9 & 23.63 $\pm$ 0.08 & 0.49 $\pm$ 0.08\\
\hline	
\end{tabular}
\begin{minipage}{0.8\textwidth}
Notes: $^{\text{a}}$ W and LP refer to wide-band and long-pass filters respectively;
$^{\text{b}}$ The uncertainties on the magnitudes are weighted by the errors on the aperture correction
$^{\text{c}}$ These magnitudes have not been corrected for extinction
\end{minipage}
\label{Tab:photometry}
\end{table*}

The field containing the counterpart to the ULX was observed during two separate HST visits with the WFC3 and ACS detectors in seven independent filters. We summarise the details of these observations in Table~\ref{Tab:photometry}. We first ran the {\tt{calwf3}} pipeline (version 3.3\footnote{https://hst-crds.stsci.edu/}) on the raw event files of all WFC3 observations using the latest calibration files available in the STScI website\footnote{The equivalent pipeline {\tt{calacs}} was used to reduce the ACS data}. The pipeline produces event files that have been bias subtracted, flat fielded and corrected for the effect of charge transfer inefficiency (CTI). We ensured that all sub-exposures were aligned to within 0.1" accuracy (using the {\tt{drizzlepac}} tool {\tt{TweakReg}}v1.4). This is especially important for the near-infrared (NIR) image, in which the ULX counterpart resides in a crowded environment, such that any misalignments between individual sub-exposures could compromise the photometric accuracy. We then implemented the PSF-fitting technique to derive instrumental magnitudes for the ULX counterpart in all six WFC3 filters using the {\tt{DOLPHOT}}v1.2 software (\citealt{r32}). This method identifies point sources above a preset detection threshold from a stack of flat-fielded and cosmic-ray cleaned images. It then uses the {\tt{TinyTim}} PSF models (\citealt{r33}; \citealt{r34}) to simultaneously fit the PSF of each source and the local background. This approach is preferred over aperture photometry in scenarios where the crowding is severe (e.g. in the near infra-red waveband), which causes the PSFs of closely spaced point sources to overlap. The software performs all the necessary pre-processing steps prior to running the photometry routine (including the masking of bad pixels affected by cosmic rays and image combination). The instrumental magnitudes were initially measured using a circular aperture of 3 pixel radius, and converted to an equivalent 10 pixel (or 0.4") magnitude by applying aperture corrections. To do this, we selected relatively bright and well isolated stars in the field-of-view, with signal-to-noise ratios larger than that of the ULX counterpart. The vertical axis intercept of a linear fit to the 3 pixel versus 10 pixel magnitude of the selected objects yields the aperture correction $\delta$. Here, we emphasise that a fitting function accounting for uncertainties in both the independent and dependent variables (equivalent to the {\tt{fit\_exy}} routine in IDL; see also \citealt{r35}) was used to ensure that the errors in $\delta$ are not underestimated. 

The initial reduction steps were similar for the ACS/SBC data.  We used the few bright and well resolved stars in the field of view to align the drizzled SBC image with the F225W and F336W filters, to confirm that the brightest object in the F140LP image is the ULX counterpart.  However, we were unable to perform PSF fitting on the ACS/SBC data due to the lack of both a model ACS/SBC PSF in {\tt{DOLPHOT}}, and of a sufficient number of well resolved stars in the field of view.  Hence aperture photometry was used within an aperture of radius $0.4$ arcseconds (equivalent to the 80\% encircled energy for ACS/SBC).  We were also hampered by the lack of bright resolved stars in calculating an aperture correction, hence we resorted to using that provided by the instrument manual \footnote{http://www.stsci.edu/hst/acs/documents/isrs/isr1605.pdf}. The results of the photometry are quoted in Table~\ref{Tab:photometry}, and we show a cleaned, 3-colour image of the field that highlights the position of the ULX counterpart in Figure~\ref{image}.

\begin{figure*}
\includegraphics[scale=0.7]{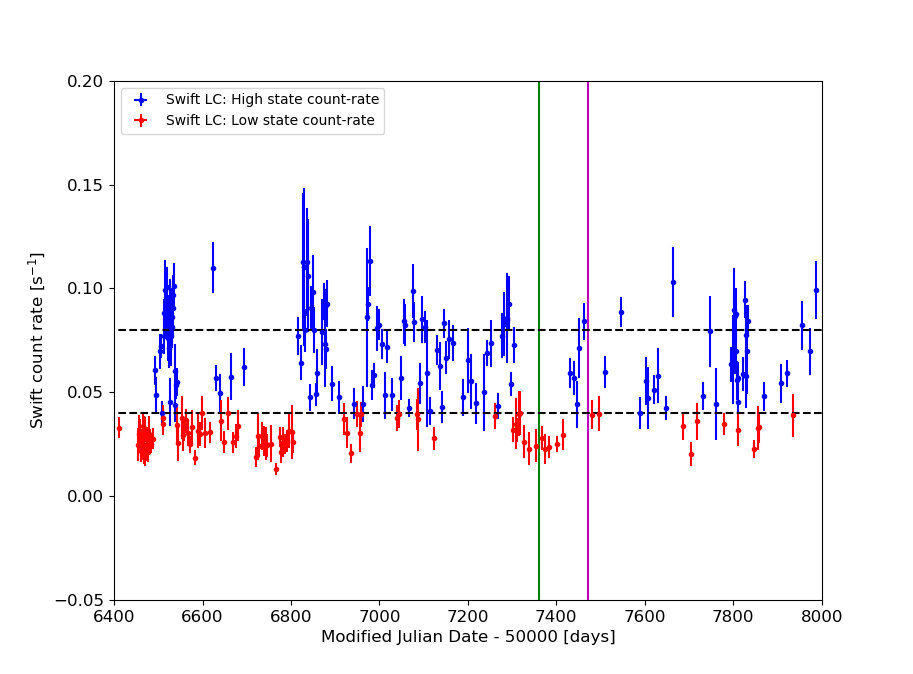}
	\caption{The long-term {\it{Swift}} light-curve of X-2, with the vertical axis plotting the 0.3-10\,keV count rate. The source appears to switch between periods of two distinct behaviours on timescales of several weeks, a relatively low flux regime (which we illustrate with red points), and a higher mean flux regime (blue points) with higher amplitude variability. The dashed black lines correspond to the two peaks in the histogram of Figure~\ref{Fig:Swiftlc2}, with the lower one used to demarcate the low and high flux regimes as displayed in the figure, while the green and magenta lines indicate the triggering of each epoch of the {\it{XMM-Newton}} and {\it{HST}} observations.}      
\label{Fig:Swiftlc1}
\end{figure*}

\begin{figure*}
\includegraphics[scale=0.5]{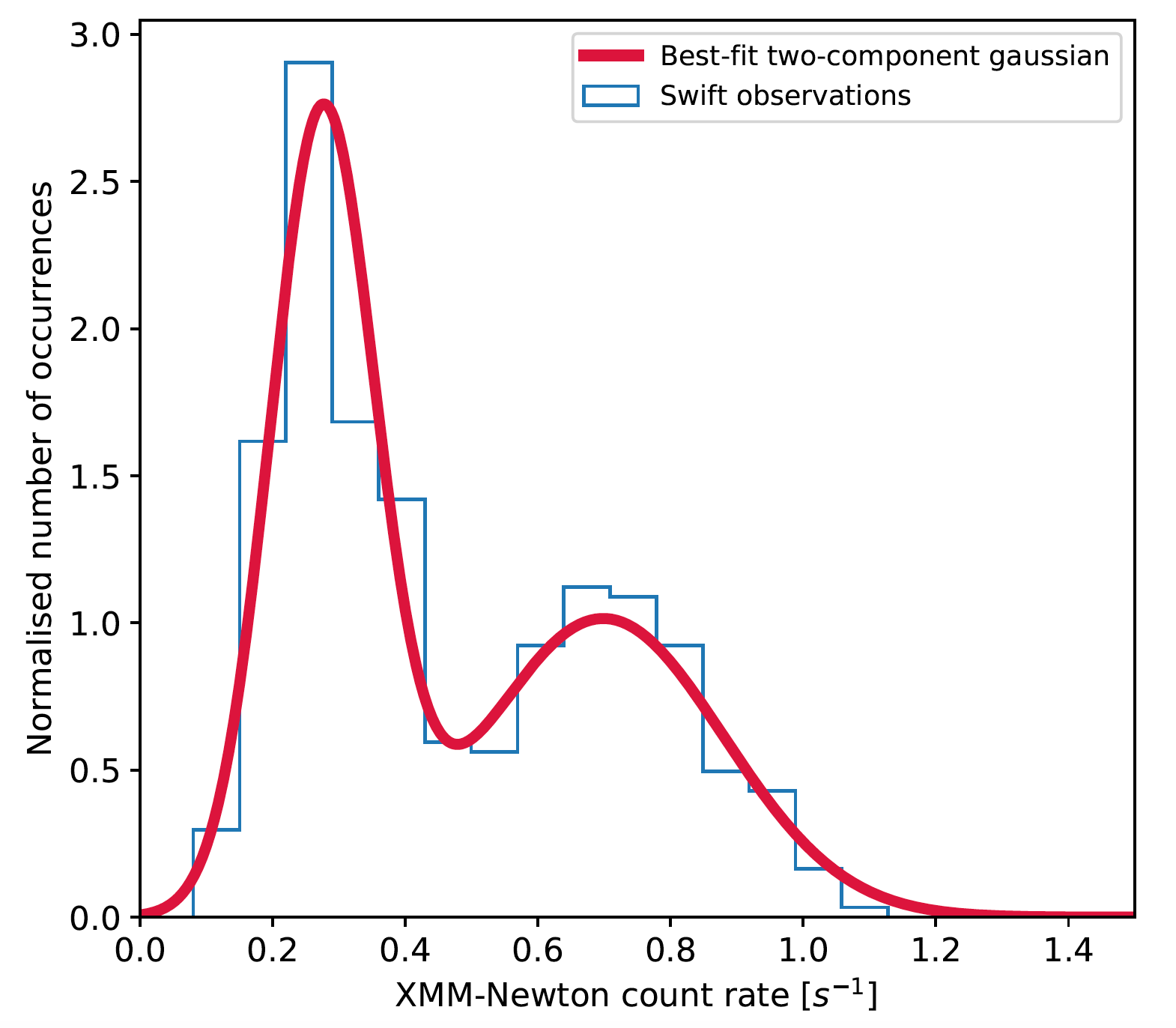}
\caption{An illustration of the bi-modality in the source X-ray flux across the entire {\it{Swift}} monitoring period of X-2. The horizontal axis specifies the count rate scaled to the {\it{XMM-Newton}} EPIC detectors combined.}      
\label{Fig:Swiftlc2}
\end{figure*}

\section{Spectral analysis}
\label{section:Spectra}
NGC 1313 X-2 was regularly monitored by the Neil Gehrels {\it{Swift}} Observatory over a baseline of several years (i.e. between April 2013 to September 2017). This has allowed us to construct its long-term X-ray lightcurve (see Figure~\ref{Fig:Swiftlc1}) using an online tool\footnote{https://www.swift.ac.uk/user\_objects/index.php} (\citealt{r1c}; \citealt{r1d}), which suggests a bimodality in the flux distribution (see Figure~\ref{Fig:Swiftlc2} and \citealt{r21}). The data allows us to visually define two distinct behaviour regimes: in the first the ULX has a relatively low count rate and displays relatively low amplitude variability on timescales of $\sim$ days. In the second, the source shows much higher amplitude variability over a range of timescales from $\sim$ weeks to days, while its mean flux level also increases.  In the second of these regimes, there is marginal evidence for a change in the hardness ratio when the ULX is at its brightest (see Figure 6 of \citealt{r21}), which becomes more prominent when the count rates are corrected for absorption. To verify these visual trends, we set a count-rate threshold to separate the low flux and high flux regimes as indicated in Figure~\ref{Fig:Swiftlc1}, and we computed the standard deviation for each of these regimes separately. Indeed, it was found that the standard deviation for the high flux regime is larger than for the low flux regime by a factor $\sim$ 4, indicating greater variability with increasing flux. The high flux regime also shows greater fractional variability (as defined in \citealt{r1e}) with $F_{\text{var}} = 0.19 \pm 0.02$, while for the low flux regime $F_{\text{var}}$ is zero (i.e. no variability in excess of that expected from noise). 

One set of our {\it{XMM-Newton}} and {\it{HST}} observations was taken in each of these behaviour regimes.  Our first observation was taken when the source count rate was relatively low and stable (green solid line in Figure~\ref{Fig:Swiftlc1}); we then followed up with our second set of observations when the source had begun to show more variable behaviour, with a higher mean count rate (magenta solid line in Figure~\ref{Fig:Swiftlc1}).

\subsection{X-ray characterisation}
\label{subsection:Xrayonly}

\begin{table*}
\caption{The best-fit parameters of the two component power-law plus multi-colour disc model, fitted to the two {\it{XMM-Newton}} spectra of X-2.}
\centering
\begin{tabular}{c c c c c c c c}
\hline
\hline
Date & $n_{\text{H}}^{\text{a}}$ & $kT_{\text{in}}^{\text{b}}$ & $\Gamma_{\text{pl}}^{\text{c}}$ & $\log_{10} L_{\text{disc}}^{\text{d}}$ & $\log_{10} L_{\text{power}}^{\text{e}}$ & Spectral regime & $\chi^{2}$/Dof$^{\text{f}}$ \\[0.5ex]
[dd-mm-yy] & [10$^{22}$ cm$^{-2}$] & [keV] & & [erg s$^{-1}$] & [erg s$^{-1}$] & &   \\
\hline
05-12-15 & 0.20$^{+0.03}_{-0.02}$ & 0.28 $\pm$ 0.03 & 1.7 $\pm$ 0.1 & 39.20 $\pm$ 0.01 & 39.69 $\pm$ 0.01 & HUL & 282.1/311 \\
24-03-16 & 0.26 $\pm$ 0.05 & 0.8$^{+0.2}_{-0.1}$ & 2.3 $\pm$ 0.3 & 39.51 $\pm$ 0.02 & 39.86 $\pm$ 0.01 & Broadened disc & 235.4/212 \\
\hline	
\end{tabular}
\begin{minipage}{0.8\textwidth}
Notes: $^{\text{a}}$ Hydrogen column density; $^{\text{b}}$ Inner disc temperature; $^{\text{c}}$ Photon index; $^{\text{d}}$ Disc component luminosity (0.3 - 10 keV); $^{\text{e}}$ Power-law component luminosity (0.3 - 10 keV); $^{\text{f}}$ Reduced $\chi^{2}$ of the fit 
\end{minipage}
\label{Tab:Fits1}
\end{table*}

\begin{figure*}
\includegraphics[scale=0.7]{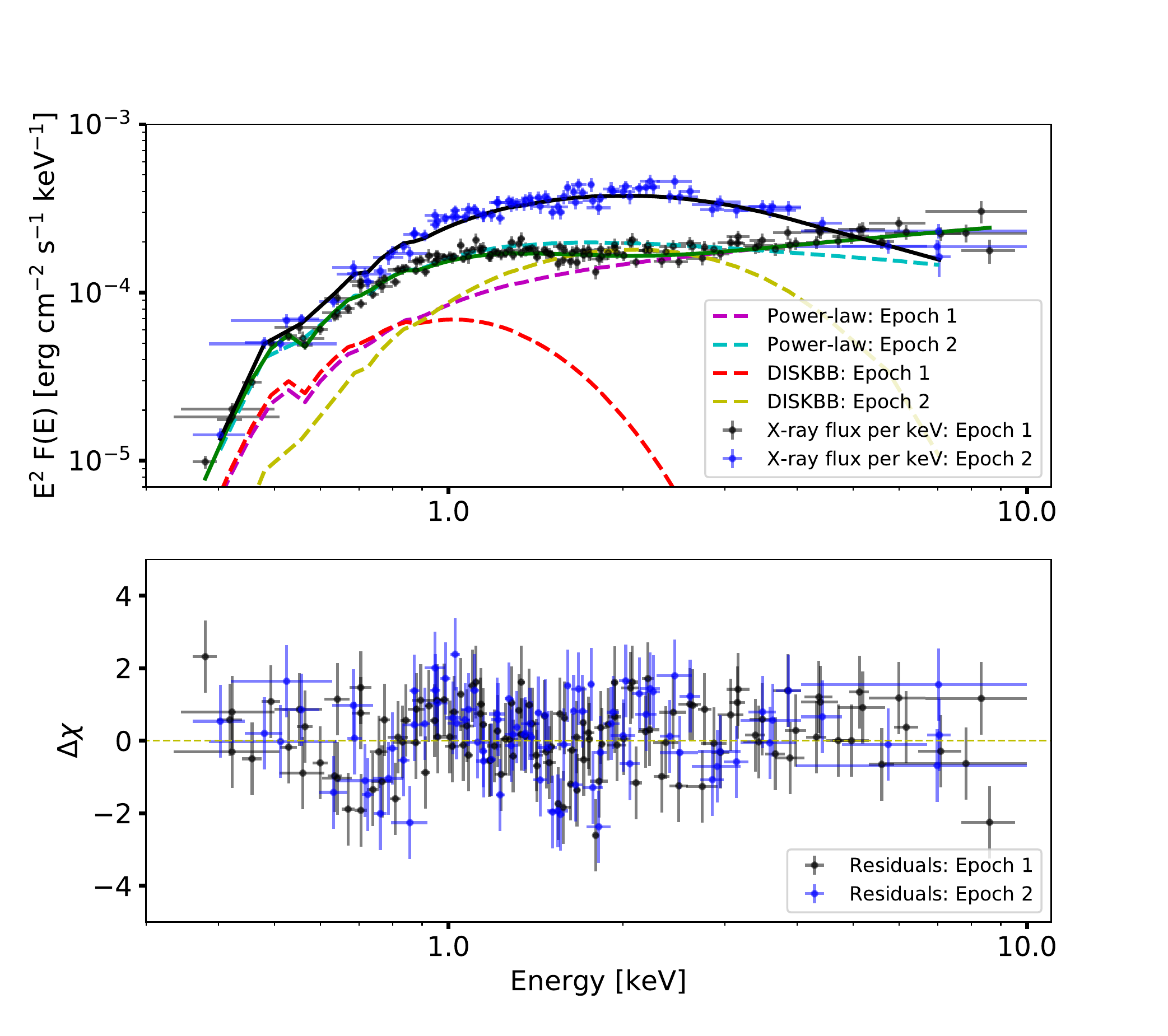}
\caption{The {\it{XMM-Newton}} X-ray spectra of X-2 with the observed data on December 2015 (black points) and March 2016 (blue points). The best-fitting multi-colour disc and power-law model to each spectrum are overlaid, with the legend specifying the label to each component. The residuals are shown in the bottom panel, indicating a good fit to the data.  The spectral shape changes between the observations}     
\label{Fig:SED2}
\end{figure*}

We began by fitting the \textit{XMM-Newton} spectra of both observations separately.   In order to obtain an understanding of the spectral shape in each observation, and any changes between observations, we fitted the data using a simple model used previously to characterise ULX spectra, i.e. an absorbed multicolour disc ({\tt{diskbb}}) plus power-law model in {\tt{xspec}}. X-ray absorption was modelled using two instances of the Tuebingen-Boulder ISM component ({\tt{tbabs}}; \citealt{r36}).  We fixed the first component to the Galactic extinction value in the direction of the source (3.97 $\times 10^{20}$ cm$^{2}$) using the {\tt{heasarc}} tool N$_{\text{H}}$\footnote{https://heasarc.gsfc.nasa.gov/cgi-bin/Tools/w3nh/w3nh.pl}, and allowed the second to vary. We added cross normalisation factors to account for any differences in calibration between the EPIC and MOS detectors, but these were found to be consistent to within 5 percent. The model provides a good fit to the data as indicated in Table~\ref{Tab:Fits1}. As can be seen in Figure~\ref{Fig:SED2} the ratios of the power-law and disc fluxes, in addition to the disc temperatures, suggest that the source spectrum changes from being in the hard ultraluminous (HUL) regime (in the first observation) to a broadened disc (BD) spectrum (see \citealp*{r14}). As discussed in that work, one caveat regarding this spectral classification is that the HUL spectra of some ULXs with pronounced curvature can be difficult to distinguish from BD spectra.  While the latter were predominantly found near the Eddington threshold in \citet{r14}, their X-ray luminosity was never part of the identification for the regime, and indeed the realisation that some ULXs have spectra that evolve from HUL to BD-like with increasing luminosity occurred subsequent to that work (e.g. see \citealt{r22}). This appears to also be the case for NGC 1313 X-2, so in what follows we use the term broadened disc to describe the shape of the X-ray spectrum while noting that this does not directly infer any physical meaning.

Critically, the observed change in spectrum supports the notion that there are real physical changes in the ULX between the low, stable 0.3-10\,keV flux behaviour in December 2015 and the higher flux, more variable periods in March 2016. 



\subsection{Does the optical emission vary?}
\label{subsection:NIRvar}

\begin{table*}
\caption{UV/optical/IR ({\it HST\/}) and X-ray ({\it{XMM-Newton}}) fluxes of X-$2$ as measured in December 2015 and March 2016.}
\centering
\begin{tabular}{c c c c}
\hline
\hline
\multicolumn{4}{|c|}{HST fluxes}\\
\hline
Pivot wavelength & Source flux (epoch 1) & Source flux (epoch 2) & Ratio$^{\text{a}}$\\[0.5ex]
[Angstrom] & [10$^{-18}$ erg s$^{-1}$ cm$^{-2}$ \AA$^{-1}$] & [10$^{-18}$ erg s$^{-1}$ cm$^{-2}$ \AA$^{-1}$] &\\
\hline
1528.0 & 38.0 $\pm$ 2.0 & 35.0 $\pm$ 2.0 & 0.92 $\pm$ 0.07\\
2375.0 & 12.0 $\pm$ 1.0 & 13.3 $\pm$ 0.6 & 1.1 $\pm$ 0.1\\
3356.0 & 5.5 $\pm$ 0.4 & 5.6 $\pm$ 0.2 & 1.0 $\pm$ 0.1\\
4327.0 & 3.1 $\pm$ 0.2 & 2.8 $\pm$ 0.1 & 0.9 $\pm$ 0.1\\
5307.0 & 1.7 $\pm$ 0.1 & 1.57 $\pm$ 0.04 & 0.92 $\pm$ 0.06\\
8048.0 & 0.47 $\pm$ 0.02 & 0.42 $\pm$ 0.02 & 0.89 $\pm$ 0.06\\
12486.0 & 0.15 $\pm$ 0.01 & 0.11 $\pm$ 0.01 & 0.73 $\pm$ 0.08\\
\hline
\hline
\multicolumn{4}{|c|}{{\it{XMM-Newton}} fluxes}\\
\hline
X-ray energy range & Source flux (epoch low) & Source flux (epoch high) & Ratio$^{\text{a}}$\\[0.5ex]
[keV] & [10$^{-13}$ erg s$^{-1}$ cm$^{-2}$ \AA$^{-1}$] & [10$^{-13}$ erg s$^{-1}$ cm$^{-2}$ \AA$^{-1}$] &\\
\hline
0.3 - 1.0 & 3.8 $\pm$ 0.1 & 4.0 $\pm$ 0.1 & 1.05 $\pm$ 0.04\\
1.0 - 4.0 & 4.2 $\pm$ 0.1 & 8.1 $\pm$ 0.2 & 1.90 $\pm$ 0.1\\
4.0 - 10.0 & 3.1 $\pm$ 0.1 & 2.4 $\pm$ 0.1 & 0.77 $\pm$ 0.04\\
\hline	
\end{tabular}
\begin{minipage}{0.8\textwidth}
Notes: $^{\text{a}}$ Ratio of source flux between the two epochs	
\end{minipage}
\label{Tab:Irrtable}
\end{table*}

\begin{figure*}
\includegraphics[scale=0.7]{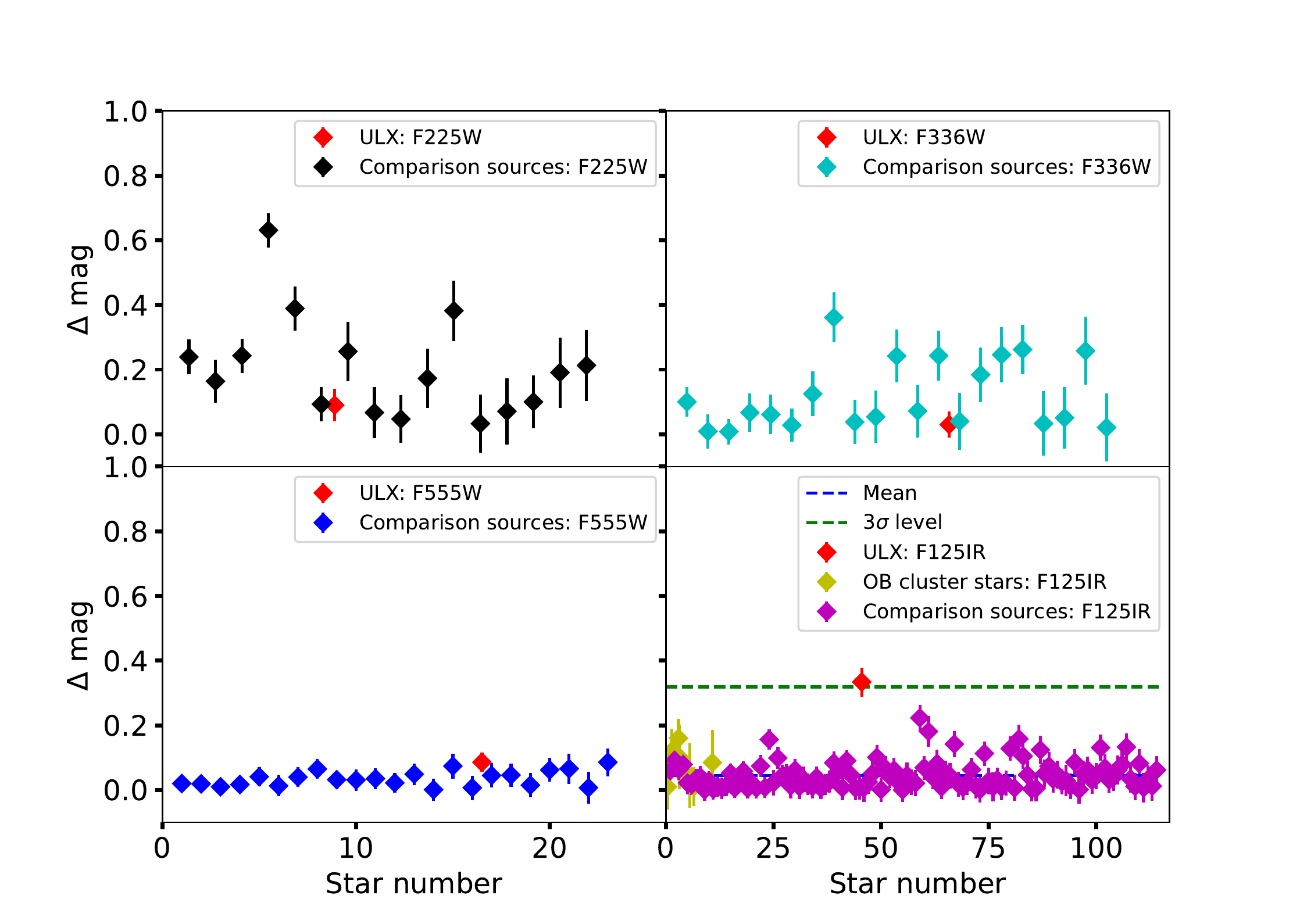}
\caption{{\it{Left}}: The observed change in the UV/optical/NIR magnitudes ($\Delta$ mag) of the ULX counterpart, selected comparison sources and stars belonging to the OB association, between December 2015 and March 2016. The particular {\it HST\/} filter in each case is indicated by the legend. The mean $\Delta$ mag of all the reference sources (dashed blue) and the corresponding 3$\sigma$ scatter (dashed green) are shown only for the NIR filter where the value of $\Delta$ mag of the ULX differs considerably from the reference sources. To re-iterate, the level of variability of the ULX is clearly consistent with most invariant stars in the field of view for all except the NIR waveband.}  
\label{Fig:NIR_variability1}
\end{figure*}

\begin{figure*}
\includegraphics[scale=0.7]{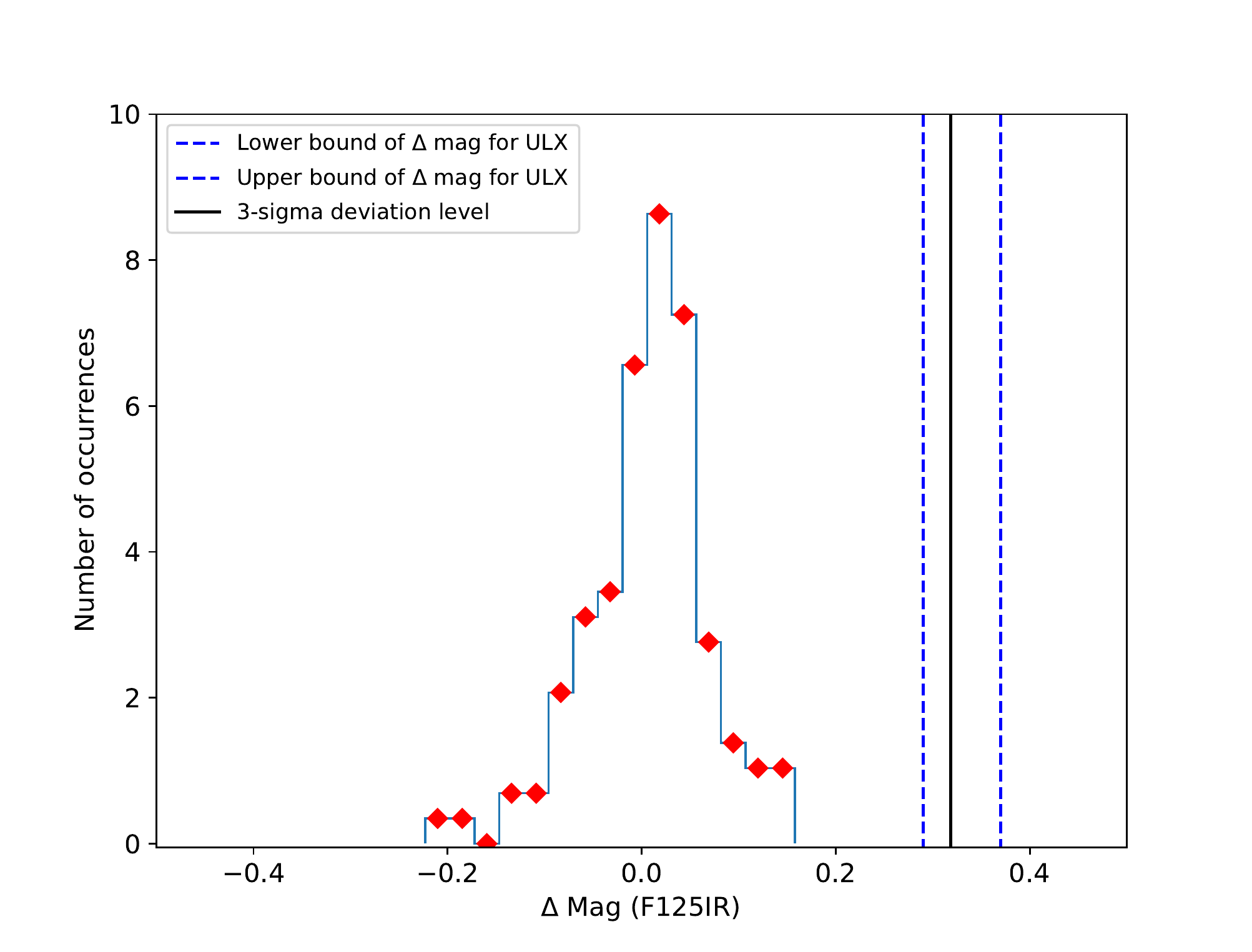}
\caption{A distribution of the NIR differential magnitudes ($\Delta$ mag) of the reference stars (red diamonds). This distribution was fitted with a Gaussian to the compute the 3-sigma deviation level (black solid line) from the mean. The $\Delta$ mag value for the ULX and its 1-sigma error region are indicated within the dashed blue lines, which clearly overlaps and exceeds the 3-sigma level.}  
\label{Fig:NIR_variability2}
\end{figure*}

In Section 3.1 we demonstrated that the X-ray emission varies between the two observational epochs (i.e. over a three month period).  The {\it HST\/} data were obtained simultaneously to the X-ray data in the two different epochs specifically to test how the UV/optical/IR colours change with the X-ray emission.  We present the observed fluxes in each filter at these epochs (converted from the magnitudes given in Table~\ref{Tab:photometry}) in Table~\ref{Tab:Irrtable}, alongside the ratio of the fluxes between the two observations.  We also present the X-ray flux separated into three bands; this again demonstrates the variability of the X-ray emission, with the largest variations occurring at medium X-ray energies (1-4\,keV).

There appears to be no statistically significant variation exceeding the 3$\sigma$ error range between the two epochs in the UV/optical bands.  However, there is some evidence for variability at longer wavelengths, particularly in the NIR (F125W) band, where the source dims by about $25\%$ ($0.33 \pm 0.04$ mags) as the X-ray emission rises from epoch 1 to epoch 2.  To validate this result, we carefully cross-matched relatively bright and well-isolated sources in both observations in the WFC/IR filter images, and in three other filters. We rejected objects that were too saturated, or those with too many bad pixels (due to cosmic ray hits), by applying a flag threshold within {\tt{dolphot}}. Aside from the ULX, our final selection included a number of stars belonging to the OB association, which should be intrinsically invariant on monthly timescales\footnote{Although we note that B stars are sometimes surrounded by circumstellar excretion discs due to their fast rotation, which can enhance variability in the infrared.}.  In any case, the bottom right panel of Figure~\ref{Fig:NIR_variability1} shows that the ULX counterpart has the largest change in the NIR magnitude (defined as $\Delta$ mag) between the two epochs, when compared to the reference sources; its $\Delta$ mag value is clearly an outlier, and deviates from the mean NIR variability of the sample by about 3 $\sigma$ (see Figure~\ref{Fig:NIR_variability2}).  This hints towards a possible first detection of NIR variability from this object, which had only been observed at UV/optical wavelengths prior to this work. We will elaborate on the possible origin of the NIR emission in the discussion.

\subsection{Optical colours \& stellar models}
\label{subsection:stellarfits}

\begin{figure*}
\includegraphics[scale=0.65]{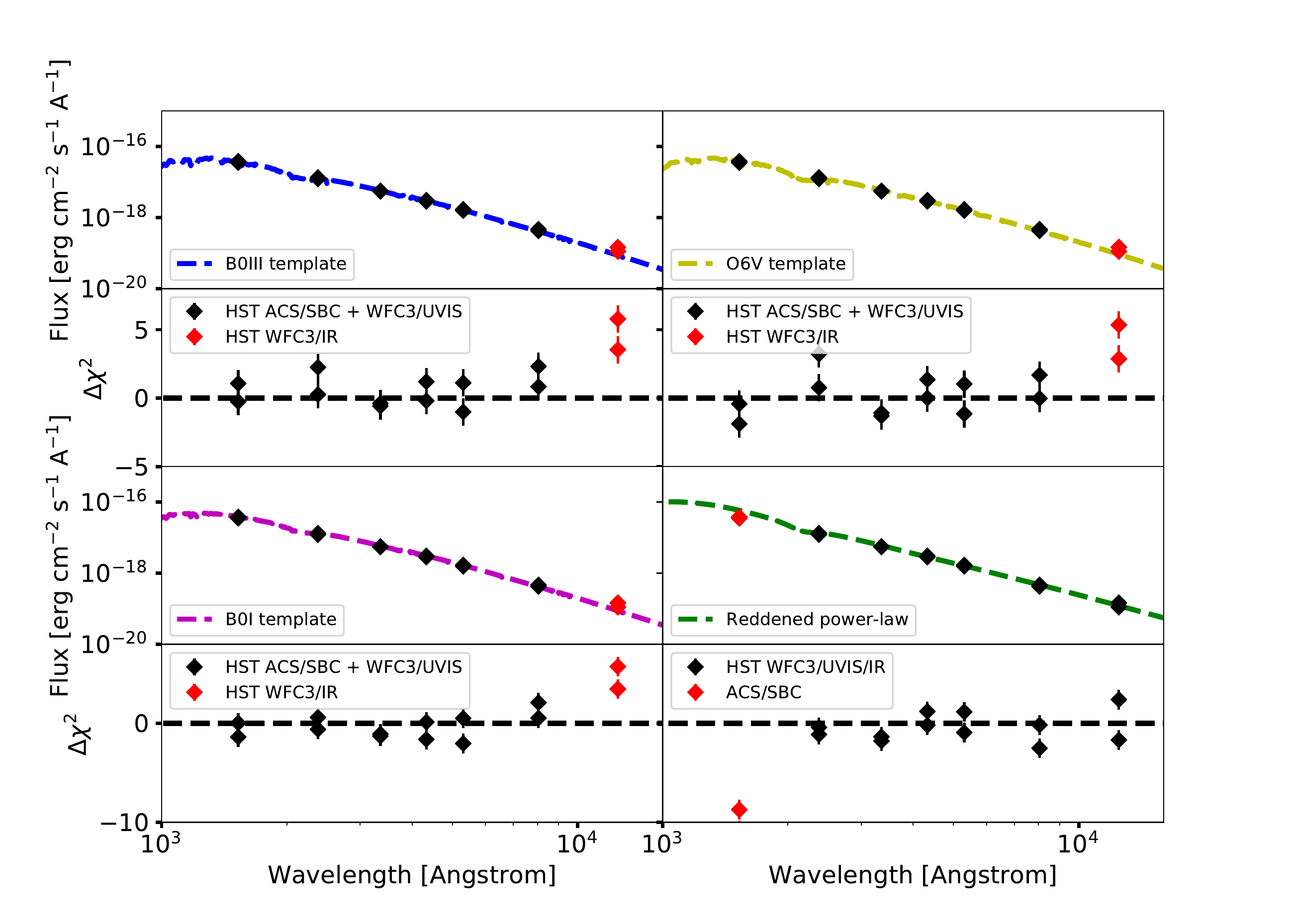}
\caption{A collection of O/B stellar templates and a reddened power-law model fit (dashed lines) to the UV/optical/NIR photometry (black diamonds) of the ULX. In all cases, the data points discrepant with respect to the best-fitting model are highlighted by the red diamond. A slight NIR excess is seen in all the stellar fits, whereas the reddened power-law model over-predicts the far UV emission.}      
\label{Fig:stellar_temp}
\end{figure*}

\begin{table*}
\caption{A collection of stellar templates and a reddened power-law model fit to the UV/optical/NIR spectrum of X-2. 
}
\centering
\begin{tabular}{c c c c c c c c c}
\hline
\hline
Model & E(B-V)$^{\text{a}}$ & T$_{\text{eff}}$ $^{\text{b}}$ & $\log$ g$^{\text{c}}$ & R$_{*}$ $^{\text{d}}$ & $\log$ Z$^{\text{e}}$ & $\Gamma$ $^{\text{f}}$ & power-law norm$^{\text{g}}$ & $\chi^{2}$/dof \\
& & [K] & & [R$_{\odot}$] & & & [1$\times 10^{-8}$] & \\
\hline
B0III$^{\text{h}}$  & 0.14 $\pm$ 0.01 & 29000$^{\dagger}$ & 3.5$^{\dagger}$ & 12.4 $\pm$ 0.3 & -0.5 & - & - & 14.5/9.0 \\
O6V$^{\text{h}}$  & 0.19 $\pm$ 0.01 & 38867$^{\dagger}$ & 4.0$^{\dagger}$ & 10.5 $\pm$ 0.2 & 0.5 & - & - & 24.6/9.0 \\
B0Ia$^{\text{h}}$  & 0.11 $\pm$ 0.01 & 26000$^{\dagger}$ & 3.0$^{\dagger}$ & 13.4 $\pm$ 0.3 & -1.5 & - & - & 13.9/9.0 \\
Power-law$^{\text{i}}$  & 0.12 $\pm$ 0.04 & - & - & - & - & 3.3 $\pm$ 0.2 & 5.0 $\pm$ 0.8 & 11.3/9.0 \\
\hline	
\end{tabular}
\begin{minipage}{0.8\textwidth}
Notes: $^{\text{a}}$ Dust reddening; $^{\text{b}}$ Effective temperature (from Martins, Scharer \& Hiller 2005); $^{\text{c}}$ Surface gravity (from Martins, Scharer \& Hiller 2005); $^{\text{d}}$ Stellar radius; $^{\text{e}}$ Metallicity (these are values are for the best fitting model from a grid of five trial metallicities per stellar type); $^{\text{f}}$ Exponent of power-law fit; $^{\text{g}}$ Power-law normalisation; $^{\text{h}}$ Obtained with the NIR data points excluded; $^{\text{i}}$ Obtained with the FUV data points excluded; $\dagger$ Parameter fixed. 
\end{minipage}
\label{Tab:Fits1b}
\end{table*}

Given that the optical emission appears to be relatively invariant between the two epochs, we examine whether the UV/optical emission of the source is consistent with arising solely from the donor star. The ULX counterpart is thought to be associated with a young OB star cluster in its vicinity, since it has a similarly large UV/optical luminosity and a distinctly blue colour, as per other members of the cluster (cf. Figure~\ref{image}). By fitting stellar isochrones to the colour magnitude diagrams (CMDs) of its member stars, \citet{r27} inferred the cluster age to be $20 \pm 5$ Myrs, implying that the mass of the ULX donor is $< 16 M_{\odot}$. Given this constraint on the counterpart, in addition to its absolute visual magnitude ($M_{\text{V}} \sim -5$) and colour index ($B-V \sim -0.25$), a range of possible OB spectral types were suggested in the work of \citet{r27}. Here, we attempt to refine these constraints by re-fitting the {\it HST\/} magnitudes (see Table~\ref{Tab:photometry}) with a grid of O/B type stellar templates from \citet{r38}. The temperature of individual spectral types were adopted from \citet*{r39}\footnote{for O-type stars} and \citet*{r40}\footnote{for all other spectral types}. The normalisation of each template depends on the stellar radius and the source distance, where we allowed the former to vary and fixed the latter to a value of 4.0 Mpc (based on a weighted average of the distances to NGC 1313 in the SIMBAD database \footnote{http://simbad.u-strasbg.fr/simbad/}). We included two additional free parameters, namely the metallic abundance and interstellar reddening (using the \citealt{r37b} extinction curve).  We fit the data from both epochs simultaneously given that we have seen little or no change in the flux between epochs in most bands.

The best fitting stellar templates are shown in Figure~\ref{Fig:stellar_temp}, and the fits are summarised in Table~\ref{Tab:Fits1b}.  A B0III giant provides the most reasonable fit (i.e. $\chi^{2}$ = 14.5 for 9 degrees of freedom), but only if the NIR data points are excluded, with the fit otherwise worsening by $\Delta \chi^{2} = 20$ for 2 degrees of freedom. A B0Ia supergiant may have a better reduced $\chi^{2}$, but the best-fit stellar radius is at least a factor of 2.5 smaller than expected for a star of this type (\citealt*{r40}).   A main-sequence O6V star can provide a reasonable fit, but it is a worse fit to the data than B-star models with $\chi^2 = 24.9$ for 9 degrees of freedom.  

We also attempted a reddened power-law model as a fit to the data.  It strongly over-predicts the far UV flux, but provides an acceptable fit at larger wavelengths.  This is intriguing for interpreting the optical emission, as reprocessed emission will have a power-law-like spectrum at optical wavelengths.  Hence it is clear from the optical data alone we cannot unambiguously identify the physical origin of the counterpart's emission.

\subsection{Irradiation of the outer accretion disc}
\label{subsection:irradiation}

Despite invariance of the optical emission in the current data set, \citet{r27} used archival VLT observations to report that the optical counterpart to the ULX shows significant variability in the B-band (by $\sim 0.2$ mag) on timescales ranging from several hours to days, with respect to other non-variable stars in the field. These timescales are much faster than that at which the mass accretion rate varies at the outer disc, which occurs on the viscous timescale: 
\begin{equation}
\tau_{\text{visc}} = \alpha^{-1} \bigg (\frac{H}{R}\bigg)^{-2} t_{\text{dyn}},
\end{equation}
where $\alpha$ is the viscosity coefficient, $H$ is the disc scale-height at radius $R$ and $t_{\text{dyn}}$ is the Keplerian dynamical timescale. We can crudely estimate that $\tau_{\text{visc}}$ is of order a year assuming $\alpha \sim H/R \sim 0.1$ for a geometrically thin outer disc, whose size is determined by the orbital period of the system containing a neutron star accretor (which we assume is $\sim 2$ days; \citealt{r6}) and a B-type companion (with a mass $\sim$ 10 M$_{\odot}$). The latter is not tightly constrained, but even if it is taken to be two orders of magnitude smaller, $\tau_{\text{visc}}$ remains much larger than the B-band flickering timescale.  Variation of the outer disc itself cannot therefore explain the previous reports of optical variability.

A possible alternative to explain this is via a disc reprocessing model: variable X-rays from the inner accretion flow can illuminate the outermost regions of the disc (\citealt*{r42}), where a fraction of the incident X-rays are absorbed and heat the underlying disc material. The highest energy photons penetrate deeper into the disc atmosphere and can thermalise with the surrounding material, re-emerging with a blackbody continuum peaking in the UV/optical band (depending on the local temperature). In this scenario, the longer wavelength emission should vary coherently with the inner X-ray emission across a broad range of timescales, although the variability will be suppressed on the shortest timescales due to a finite propagation time between the two regions. In particular, it will be limited by the light travel time in the simple picture of X-rays directly illuminating a geometrically thin outer disc. If the disc is tidally truncated with an outer radius $R_{\text{out}}$ of $\sim$ 0.7 times the radius of the neutron star's Roche lobe, the light travel time is $\lesssim$ a minute (for the orbital period estimate used above). Thus, one might expect changes in the UV/optical and X-ray fluxes to be correlated over the much longer three month (inter-observational) timescale. However, a comparison of the fluxes in the two separate bands demonstrates no evidence for such a correlation. The broadband 0.3-10.0\,keV flux increases by $\sim 30 \%$ between the two observations, while the UV/optical fluxes are mostly consistent between the two epochs and only change significantly (and in the opposite sense) for the longest wavelength filter (see Table~\ref{Tab:Irrtable} and Figure ~\ref{Fig:NIR_variability1}). Therefore, we do not observe any correlated X-ray-to-optical long-term variability in the system, at least between the two epochs at which our new observations were taken.  Nevertheless given the previous evidence we cannot rule out X-ray reprocessing as a contributor to the optical emission, and so seek to constrain its presence using further techniques.

\subsubsection{X-ray irradiation model}
\label{subsubsection:Xrayrep}

\begin{figure*}
\includegraphics[scale=0.7]{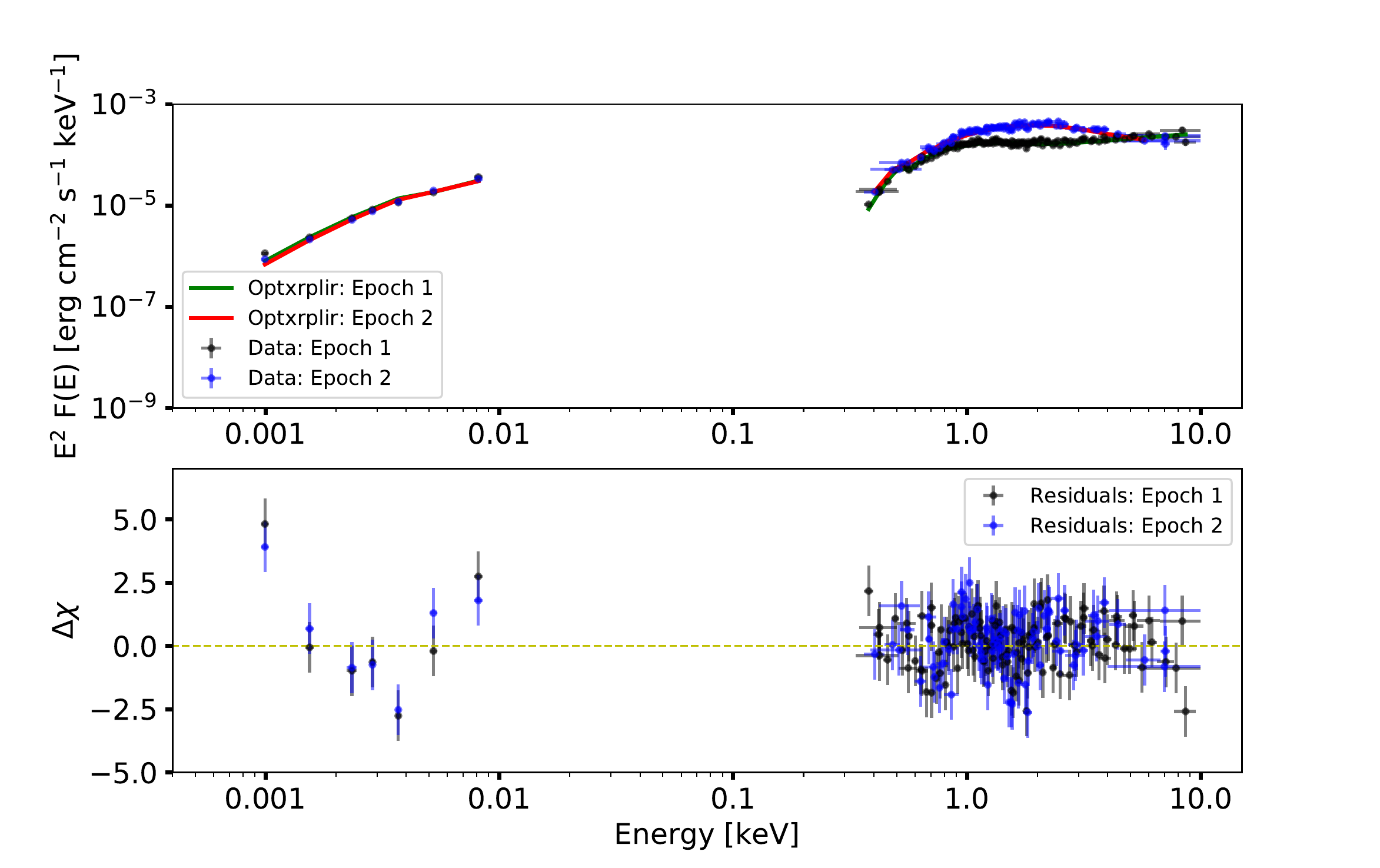}
\caption{The X-ray to optical SED of NGC 1313 X-2 taken during epoch 1 (black dots) and epoch 2 (blue dots). The broadband (0.3-10.0\,keV) X-ray flux is seen to vary by $\sim 30\%$ between the two epochs, but there is no evidence for corresponding UV/optical variability.  The best-fitting (absorbed) models are indicated by the legend. The bottom panel shows the residuals to the best-fitting models.}        
\label{Fig:SED1}
\end{figure*}

\begin{table*}
\caption{The best-fitting parameters of the irradiated disc model for both epochs.}
\centering
\begin{tabular}{c c c c c c c c c c}
\hline
\hline
Epoch & $n_{\text{H}}$$^{\text{a}}$ & E(B-V)$^{\text{b}}$ & $\log_{10} \frac{L}{L_\text{Edd}}$ $^{\text{c}}$ & f$_{\text{out}}$$^{\text{d}}$ & R$_{\text{cor}}$$^{\text{e}}$ & $\Gamma$$^{\text{f}}$ & Mass$^{\text{g}}$ & $\log R_{\text{out}}$$^{\text{h}}$ & $\chi^{2}$/Dof$^{\text{i}}$ \\[0.3ex]
& [10$^{22}$ cm$^{-2}$] & (vega mag) &  &  & [$R_{\text{g}}$] & & [M$_{\odot}$] & $[R_{\text{g}}]$ & \\[0.3ex]
\hline
1 & 0.18 $\pm$ 0.02 & 0.15 $\pm$ 0.02 & -0.03$^{+0.2}_{-0.1}$ & $>0.87$ & $72^{+11}_{-7}$ & 1.7 $\pm$ 0.1 & 49$^{+8}_{-12}$  & 5.3 $\pm 0.1$ & 573.3/522 \\
\rule{0pt}{2.5ex}    
2 & 0.14 $\pm$ 0.02 & 0.12 $\pm$ 0.02 & -0.2$^{+0.25}_{-0.1}$ & $>0.86$ & 20$^{+9}_{-3}$ & 2.5$^{+0.3}_{-0.5}$ \\
\hline
\end{tabular}
\begin{minipage}{0.8\textwidth}
Notes: $^{\text{a}}$ Hydrogen column density; $^{\text{b}}$ Dust reddening; $^{\text{c}}$ Eddington ratio; $^{\text{d}}$ Reprocessed fraction; $^{\text{e}}$ Coronal radius; $^{\text{f}}$ Photon index; $^{\text{g}}$ Calculated compact object mass (which is known to be incorrect as this ULX contains a neutron star); $^{\text{h}}$ Outer disc radius (NB -- the compact object mass and outer disc radius were held constant for both fits; other parameters were left free to fit individual spectra) ; $^{\text{i}}$ Reduced chi-squared for the simultaneous fit of both SEDs
\end{minipage}
\label{Tab:Fits2}
\end{table*}

Previous work by \citet{r45} constrained the fraction of photons emitted in the central regions of the accretion disc and reprocessed in its outer regions, for a small number of ULXs with broadened disc spectra, including NGC 1313 X-2.  They used standard reprocessing models developed for sub-Eddington accretors, on the basis of which they disregarded the details of the X-ray part of the joint X-ray to optical fits as unphysical. However, they could trust the calculated reprocessed fractions as a reliable output of the model, given the energy distribution of the irradiating X-rays was broadly correct.  We take a similar tack here.

We adopted a state-of-the-art irradiated disc model ({\tt{optxrplir}}; \citealt{r43}), which assumes that the gravitational power released by the accretion flow is dissipated via the standard Novikov-Thorne emissivity, giving rise to two distinct spectral components. This includes thermal emission from an optically thick accretion disc and a warm comptonisation component emitted by a cool, optically thick corona comprising a thermal population of electrons with optical depth $\tau$ and temperature $kT_{\text{es}}$. These components are thought to be emitted in two distinct regions of the accretion flow, with the size of the corona denoted by $R_{\text{cor}}$. We neglected the contribution from the third component, corresponding to hot comptonised emission from an optically thin corona (characterised by photon index $\Gamma_{\text{pl}}$ and electron temperature $kT_{\text{eh}}$), since it is expected to be dominant far above the energy band probed by {\it{XMM-Newton}}. All these X-ray spectral components illuminate the outer disc, producing reprocessed emission in the UV/optical band. The strength of the irradiation $f_{\text{out}}$ depends on the scale-height of the outer disc and its albedo. The former is influenced by the shape of the incident X-ray spectrum, which is self consistently accounted for by the model. We note that the reprocessed fraction is defined differently in previous irradiated disc models as compared to {\tt{optxrplir}}, which separates it into a geometry dependent factor $f_{\text{out}}$ and the albedo $\alpha$: i.e. $f_{\text{out, old}} = f_{\text{out, new}} (1 - \alpha)$. Furthermore, the shape of the reprocessed emission depends on the outer disc radius $R_{\text{out}}$; a larger $R_{\text{out}}$ allows the emission to thermalise over a larger area, thus decreasing the outer disc temperature and increasing the luminosity at longer wavelengths. 

Additional parameters of the model include the black hole mass ${M}_{\text{BH}}$ and spin $a^{*}$, the Eddington ratio $\dot{m}_{\text{Edd}}$ and the distance to the object $d_{\text{Mpc}}$.  Since NGC 1313 X-2 is a neutron star and not a black hole we have an issue in using the above model to describe the physical reality of the system: the model cannot provide realistic solutions for the accretor mass and spin once the Eddington limit is locally exceeded in the accretion disc, as the Novikov-Thorne emissivity requires modification.  Hence the values of these parameters (and the linked Eddington ratio) are largely physically irrelevant for these fits of NGC 1313 X-2.  We fixed the black hole mass and spin values to be constant between observations, but otherwise simply allowed the parameters to find values that allow us to trace the X-ray spectrum as accurately as possible, as this affects $f_{\text{out}}$, but do not comment further on their best-fitting values.

We fixed the electron temperature and optical depth of the warm comptonisation component to arbitrary values ($kT_{\text{es}}$ = 0.2 keV and $\tau$ = 10 respectively, consistent with previous observations of ULXs cf. \citealt{r105}), since these could not be constrained with the current quality of data. However, we note that varying these parameters even over a range of three dex does not have a noticeable effect on the spectral fit. The albedo was set to 0.9 assuming a highly ionised outer disc (see \citealt{r44}; \citealt*{r42}). We treated dust extinction in the same manner as in section~\ref{subsection:stellarfits}, but adopted {\tt{phabs}} to model the gas phase absorption intrinsic to the source. The model normalisation is determined from ${M}_{\text{BH}}$, $\dot{m}_{\text{Edd}}$ and the inclination angle $i$, where the latter was set to a moderately face-on view of 30 degrees (given the detection of pulsations and the X-ray spectrum of X-2). Finally, we adopted a distance of 4.0 $\pm$ 0.8 Mpc towards the galaxy NGC 1313 and allowed the remaining parameters to vary between both observations.

We fitted this model to the X-ray to optical Spectral Energy Distributions (SEDs) of both epochs.  The fits were performed simultaneously with some parameters held to the same value at each epoch.  We detail the best fitting parameters in Table~\ref{Tab:Fits2} and show the data and best-fitting model in Figure~\ref{Fig:SED1}.  This model is formally an acceptable fit to the data, as it cannot be rejected above a significance of 2$\sigma$. However, the fit is obviously statistically dominated by the X-ray data (which itself shows some residuals, consistent with features described by \citealt{r37} and thought to be linked with absorption in an outflowing wind), and the shape of the optical spectrum is not well described as possible excesses are clear in both the IR and UV wings of the spectrum.

The inferred reprocessed fractions are unphysically large (i.e. close to unity), clearly inconsistent with the fact that both the donor star and the outer disc only cover a small fraction of the sky. They are also comparatively much larger than found previously for this object (and several other ULXs) by \citet{r45}. This is initially surprising, since neither the ULX optical fluxes nor the X-ray spectral shapes greatly differ between the two studies. However, the reprocessed fraction depends on the choice of the albedo $\alpha$ (i.e. thermalisation fraction) of the outer disc, among other things (see below). In the version of the irradiated disc model used by \citet{r45}, both $\alpha$ and $f_{\text{out}}$ are combined into a single free parameter, while here the former is fixed to a desired value. In principle, re-fitting the SED with a smaller albedo (0.3) yields a best-fit $f_{\text{out}}$ that is more consistent with \citet{r45}.However, it is unclear whether such a small albedo is physically plausible for this source, since it depends on the poorly understood radiative transfer of photons from the inner disc through the outflowing material. We note that more recent work by \citet*{r1i} argues that it is possible to describe the SED of NGC 1313 X-2 with a reprocessing model that yields a reasonable value of $f_{\text{out}} = 0.01$. However, their study uses a much narrower wavelength range compared to our work, and even so, their model does not fully fit the far UV portion of the SED. If we refit the SED of the source using the same wavelength range as \citet*{r1i}, we obtain a reprocessed fraction that is consistent with theirs after adopting the same albedo that they use (i.e. 0.7). 


In our case, it is difficult to pinpoint a single reason for the resulting large reprocessed fractions, since this depends on several factors. However, we try to outline a few points (aside from the choice of the albedo) that could explain such an occurrence. One critical reason is linked to the negligence of advection and mass loss by the irradiated disc model, which are important when the accretion rate is larger than Eddington (see next section). The model uses an emissivity applicable to a sub-Eddington geometrically thin disc, for which the luminosity at each radius scales linearly with the mass accretion rate ($L = \eta \dot{M} c^2$). For an advection dominated slim disc, this only applies to the outermost UV/optically emitting regions where advection is not important. The luminosity of the inner X-ray emitting region saturates when $\dot{M}$ exceeds a critical value, due to photon trapping and mass loss from a large scale height flow. Hence, the true mass accretion rate at the outer disc (and the UV/optical flux) can be severely underestimated if the standard linear scaling between $\dot{M}$ and $L$ is used (as this model does). This could easily over-predict the reprocessed fraction required to produce the observed UV/optical flux through X-ray irradiation alone. Further, the reprocessing geometry is not as simple in ULXs as assuming that X-rays from the inner flow directly irradiate the outer disc. The emergence of a radiatively driven wind (when the local disc flux is larger than Eddington) can influence the reprocessed fraction in at-least two distinct ways. Photons scattering off the optically thick regions of the wind (close to the spherisation radius) may lose enough energy as to not thermalise with material in the outer disc (\citealt*{r47}). However, some photons can also scatter off the optically thin phase of the wind at larger distances from the compact object (\citealt{r48}), which can enhance the scattering of the hard X-ray flux onto the outer regions. We cannot account for these complexities with the current model. But, \citet{r45} speculated that since the reprocessed fractions of several ULXs were comparable to that inferred for sub-Eddington X-ray binaries, these opposing effects could cancel out.  They do not appear to do so for this current model, so we conclude that it does not provide any physically meaningful constraints on this system and move on to seek out a model that can.

\subsection{Composite model including a slim disc} 
\label{subsection:slimdisc}

\begin{figure*}
\includegraphics[scale=0.5]{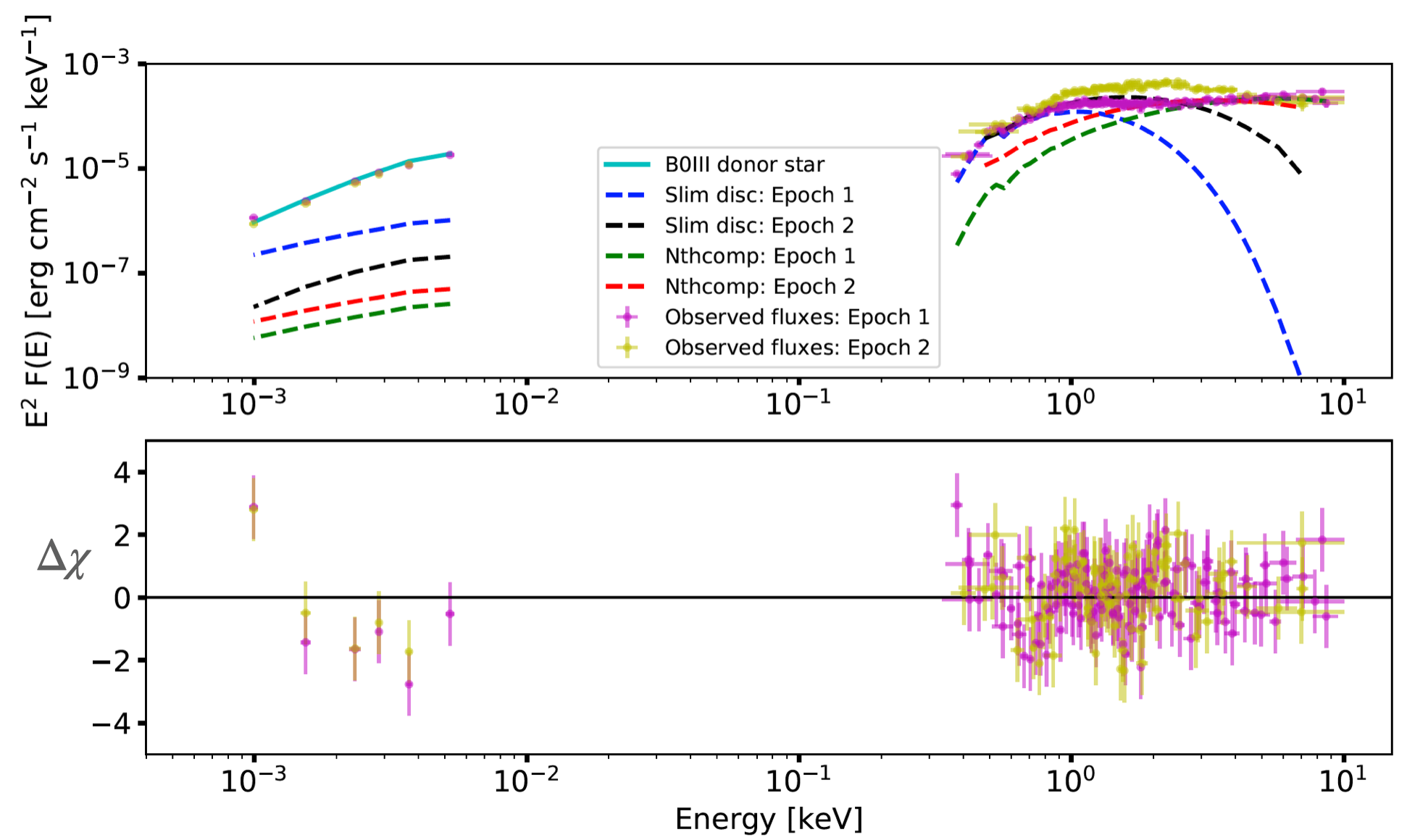}
\caption{The SED of NGC 1313 X-2 fitted with a composite slim disc and thermal comptonisation model, including the contribution from the donor star (solid cyan line). The best-fitting slim disc component is indicated by the dashed blue (epoch 1) and black (epoch 2) lines, and the corresponding thermal comptonisation component is shown by the dashed green and red lines. The residuals to the best-fitting model are shown in the bottom panel, indicating that a good fit is obtained to most of the data points.}
\label{Fig:slimdisc_with_star}
\end{figure*}

\begin{figure*}
\includegraphics[scale=0.6]{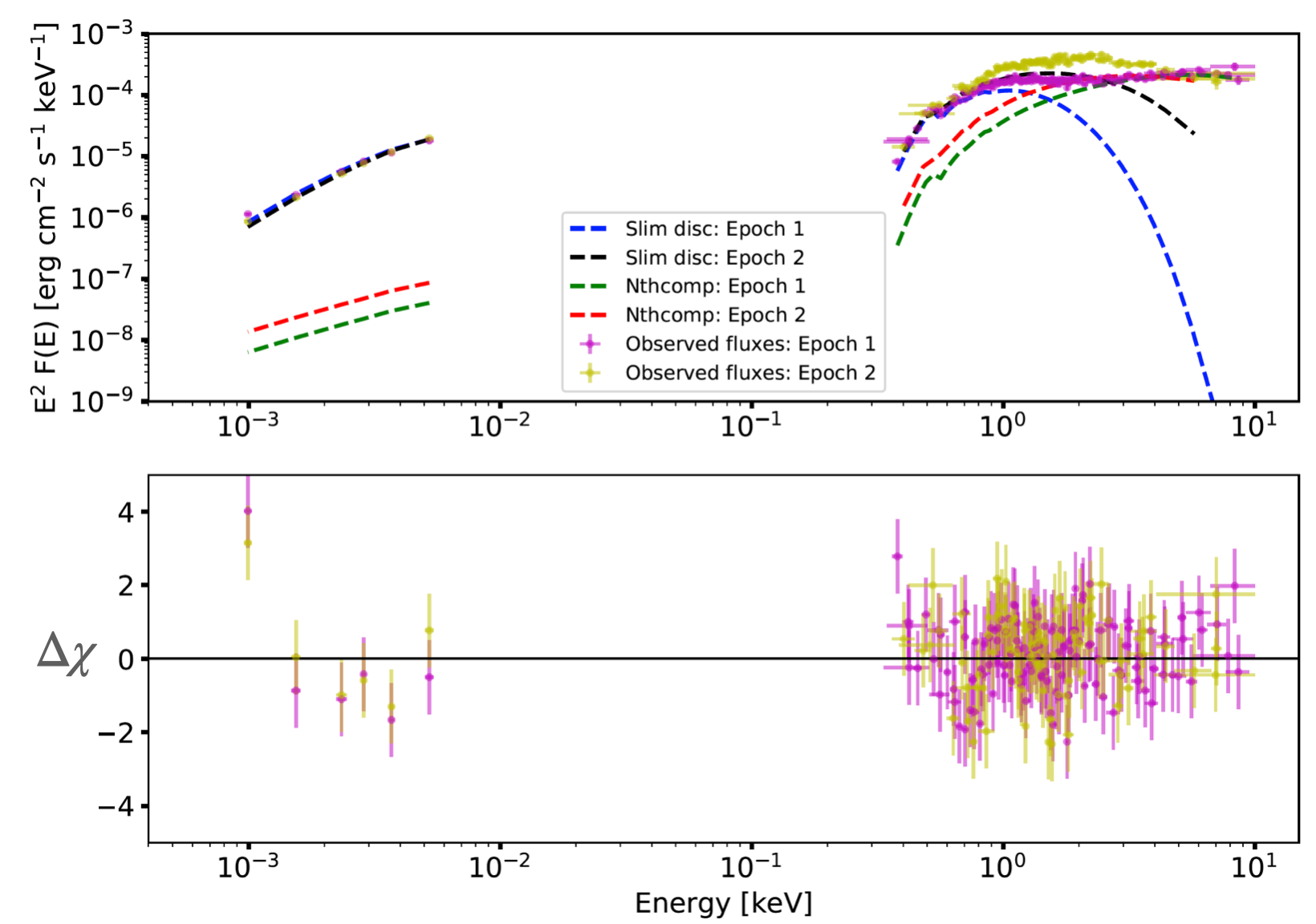}
\caption{The same plot as in Figure~\ref{Fig:slimdisc_with_star}, excluding the contribution from a donor star. The slim disc component alone provides a good fit to the UV/optical emission, although predictably this requires a substantial increase in the mass accretion rate compared to the case where the donor emission is considered.}
\label{Fig:slimdisc_no_star}
\end{figure*}

We have seen that the X-ray emission changes between epochs but that the UV/optical/IR largely doesn't; that the optical emission alone could be described by either stellar light or reprocessed X-ray emission; but that a state-of-the-art sub-Eddington irradiation model does not provide a meaningful constrain on the possible reprocessing.  We therefore attempt to construct a new model that takes into account the super-Eddington nature of the central regions and that can also account for the optical light.

Our prime means of achieving this is to replace the multi-colour disk blackbody component in the X-ray spectrum with the physically-motivated slim disc model of \citealt*{r49}, which is referred to as {\tt{agnslim}}.   In previous studies of the two-component ULX X-ray spectra, this component has been interpreted as the photospheric region at the base of the wind (\citealt{r1} and references therein) although more recent models based on magnetic neutron stars have more classically attributed it to the outer accretion disc \citep{r109}\footnote{This attribution is very problematic, given that the outer discs are assumed to be 'normal' optically-thick geometrically-thin sub-Eddington discs (which means their emission cannot be substantially beamed), yet still apparently radiate well in excess of the Eddington limit for a neutron star.}; in broadened disc spectra this component has frequently been attributed to an advection-dominated slim disc \citep{r110}. This model suppresses the luminosity released by the inner accretion flow, relative to that predicted by the accretion rate $\dot{M_{\text{in}}}$ through the outer disc. Hence, beyond some critical value of $\dot{M_{\text{in}}}$ the intrinsic X-ray luminosity of a slim disc saturates to $\sim 10$ $L_{\text{Edd}}$. The model retains the same three-radial zone structure as {\tt{optxrplir}}, with two distinct (warm and hot) coronal regions. However, a critical assumption is that the puffed-up disc at the radius $r_{\text{c}}$ self-shields the hard X-ray emission from the inner accretion flow, such that any reprocessed emission from the outer disc is negligible (see Figure 12 of Kubota \& Done 2019 for a visualisation of this effect).  


We made slight adjustments to some of the model parameters before it was appropriate for fitting the multi-wavelength spectrum of a neutron star ULX. These sources can have strong magnetic fields that impart an opposing pressure to the inward ram pressure of the infalling material, thus halting the accretion flow at a radius $R_{\text{M}}$ much larger than the innermost stable orbit $R_{\text{isco}}$. The value of $R_{\text{M}}$ depends on the dipolar field strength and the mass accretion rate at that radius (see \citealt{r50}). So, we enforced the inner disc radius of {\tt{agnslim}} to take on values much larger than $R_{\text{isco}}$ (while still allowing it to vary). Moreover, we modelled the emission from the accretion envelope surrounding the NS surface (reference) using the thermal comptonisation model ({\tt{nthcomp}}) following \citet{r53}. Further, we adopted a canonical NS mass of 1.5 M$_{\odot}$, and retained the same neutral gas absorption and dust reddening components as in the previous section. 


\begin{table*}
\caption{The best-fit parameters of the multi-component model including a slim disc. Parameters (c-f) were free to vary, while the dust reddening was linked to the value derived from \citet{r27}.}
\centering
\begin{tabular}{c c c c c c c c}
\hline
\hline
\multicolumn{8}{|c|}{With stellar contribution (using $T_{*} = 29000$ K and $R = 12.4 R_{\odot}$)}\\
\hline
Epoch & E(B-V)$^{\text{a}}$ & $n_{\text{H}}$$^{\text{b}}$ & $\log({\dot{m}})$$^{\text{c}}$ & $R_{\text{in}}$$^{\text{d}}$ & $\log R_{\text{out}}$$^{\text{e}}$ & $kT_{\text{bb}}$$^{\text{f}}$ & $\chi^{2}$/dof \\
& [Vega mag] & [10$^{22}$ cm$^{-2}$] & & [$R_{\text{g}}$] & [$R_{\text{g}}$] & [keV] & \\
\hline
1 & 0.09 $\pm$ 0.01 & 0.16 $\pm$ 0.02 & 1.22 $\pm$ 0.1 & 129 $\pm 14$ & 6.0$^{*}$ & 1.68 $\pm$ 0.2 & 283.01/310 \\
\rule{0pt}{2.5ex}    
2 & 0.12 $\pm$ 0.01 & 0.15 $\pm$ 0.02 & 1.17 $\pm$ 0.04 & 24$^{+12}_{-10}$ & 6.0$^{*}$ & 0.9 $\pm$ 0.1 & 258.4/212 \\
\hline
\hline
\multicolumn{8}{|c|}{Without stellar contribution}\\
\hline
Epoch & E(B-V)$^{\text{a}}$ & $n_{\text{H}}$$^{\text{b}}$ & $\log({\dot{m}})$$^{\text{c}}$ & $R_{\text{in}}$$^{\text{d}}$ & $\log R_{\text{out}}$$^{\text{e}}$ & $kT_{\text{bb}}$$^{\text{f}}$ & $\chi^{2}$/dof \\
& [Vega mag] & [10$^{22}$ cm$^{-2}$] & & [$R_{\text{g}}$] & [$R_{\text{g}}$] & [keV] & \\
\hline
1 & 0.16$^{+0.01}_{-0.05}$ & 0.25 $\pm$ 0.01 & $>4.3$ & 133$^{+11}_{-10}$ & 6.4$^{+0.1}_{-0.02}$ & 1.6$^{+0.1}_{-0.05}$ & 266.3/309 \\
\rule{0pt}{2.5ex}    
2 & 0.1$^{*}$ & 0.24 $\pm$ 0.01 & 3.86$^{+0.02}_{-0.04}$ & 25$^{+12}_{-11}$ & 6.6 $\pm$ 0.03 & 0.86 $\pm$ 0.1 & 253.5/211 \\
\hline
\end{tabular}
\begin{minipage}{0.8\textwidth}
Notes: $^{\text{a}}$ Dust reddening; $^{\text{b}}$ Hydrogen column density; $^{\text{c}}$ Dimensionless mass accretion rate (equivalent to the Eddington ratio); $^{\text{d}}$Inner-disc radius; $^{\text{e}}$ Outer-disc radius; $^{\text{f}}$ Temperature of seed (blackbody) photons to non-thermal electrons; $^{*}$ Parameter frozen because it is insensitive to the fit.
\end{minipage}
\label{Tab:Fits3}
\end{table*}

In Table~\ref{Tab:Fits3}, we present the best-fit parameters for the two cases where (i) we fit the multi-wavelength SED of X-2 with a composite model including {\tt{agnslim}}, {\tt{nthcomp}} and a single-temperature blackbody ({\tt{bbodyrad}}) to account for the donor star's emission, and (ii) when the latter is ignored.  Discarding any stellar contribution in the UV/optical yields an upper limit to $\dot{m}$ that we annotate as $\dot{M}_{0}$.  But, we emphasise that due to uncertainties in the donor spectral type and any other contributions to the {\it HST\/} spectrum, the latter will invariably have some unknown systematic uncertainty. We set the temperature and normalisation of {\tt{bbodyrad}} to match a B0III spectral type, since this gives the most reasonable (although not an ideal) fit to the UV/optical magnitudes in section~\ref{subsection:stellarfits}.  We illustrate the fits to the SED of NGC 1313 X-2 with this new model both with and without a stellar contribution in Figures~\ref{Fig:slimdisc_with_star} and \ref{Fig:slimdisc_no_star} respectively.

Although the models generally provide good fits across the SED, it is noticeable that the IR datapoint presents an outlier to both models.  This is because neither the stellar blackbody produces sufficient flux at these longer wavelengths to match the data, nor does the accretion disc become large enough (given the suspected orbital constraints) for its outer radii to radiate with a characteristic temperature in this band.  We must therefore seek a different explanation for the flux in this band, and do so in the Discussion section.



Table~\ref{Tab:Fits3} shows that statistically acceptable fits are obtained for both cases (i) and (ii), with the major difference being the requirement of a substantially larger $\dot{M}_{0}$ when the stellar emission is not considered (i.e. $\gtrsim 6000$ times the Eddington value). One reason to disfavour this solution is the following: for a neutron star accretor, any advected energy must be released near their surface (unlike the case of a black hole, where it falls beyond the event horizon).  In the absence of strong disc winds (as the \citealt{r55} prediction implies), we can estimate the luminosity radiated near the surface to be $\sim 5 \times 10^{41}$ erg s$^{-1}$, using the accretion efficiency applicable to a neutron star ($\eta = \frac{G M_{\text{NS}}}{R_{\text{NS}} c^{2}} \sim 0.2$) and the lower best-fit $\dot{M}_{0} \sim 6000 \dot{M}_{\text{Edd}}$.  This is clearly at odds with the much smaller observed luminosity of the thermal comptonisation component ($3 \times 10^{39}$ erg s$^{-1}$), implying that the outflow rate must be more than than 95\% of the accretion rate through the outer disc, and that implausibly strong winds are required\footnote{Estimates from ULX bubbles, including for NGC 1313 X-2, are that winds carry $\lesssim 10^{40} ~\rm erg~s^{-1}$ of kinetic energy (e.g. \citealt{r111}), over an order of magnitude smaller than required here.}.  Admittedly, we could be underestimating the luminosity from the accretion column above 10 keV, with analyses of the pulsed emission of other NS ULXs indicating that it peaks above the bandpass of {\it{XMM-Newton}}. However, this discrepancy is likely to be well within a factor of 2 (see e.g. \citealt{r56}). Therefore, the best-fit mass inflow rate inferred by only considering the slim disc contribution at UV/optical wavelengths leads to inconsistencies on the following front: the prediction of an unreasonably large luminosity from the NS surface and/or an implausibly powerful wind. We note that the best-fit $\dot{m}$ is also influenced by the inclination $i$, since the projected emission from the accretion flow scales as $\cos i$ (i.e. larger viewing angles will require an increase in $\dot{m}$ to fit a given spectrum). Setting the inclination to an extremal 0 degrees (fully face on view) does reduce the best-fit $\dot{m}$, but not to the extent that the above problems are alleviated.  Hence, we strongly favour the fits including a stellar component on a physical basis.  


The derived mass accretion rate $\dot{m}$ hardly changes between the two observations. This is expected given a lack of observed variability in the UV/optical bands, which correspond to emission from the outermost regions of the disc that are much more sensitive to changes in $\dot{m}$ (since photon trapping effects are unimportant here). Despite an unchanged mass accretion rate, the inner disc radius takes on a much smaller value in the second epoch (decreasing by a factor $\sim 5$), which allows {\tt{agnslim}} to account for an increase in the 1.0 - 6.0 keV flux.  This is difficult to understand if $R_{\text{in}}$ is interpreted as the magnetospheric radius, which scales primarily with the mass accretion rate and magnetic field strength as $\dot{M}^{-\frac{2}{7}} B^{\frac{4}{7}}$ (\citealt*{r57}). Unless the field strength decreases considerably (by a factor $\sim 10$) on the three month timescale, there is no reason to expect any variation in $R_{\text{M}}$.

\section{Discussion}
\label{section:discussion}

We have presented the results of simultaneous X-ray and UV/optical/NIR photometry of the pulsating ULX NGC 1313 X-2.  The results of our campaign show that when the X-ray emission (monitored by {\it{Swift}} and {\it{XMM-Newton}}) appears to transit from a low, stable flux to a higher, more variable flux in the 0.3-10\,keV regime, the X-ray spectral shape shows notable variability without being accompanied by any changes in the UV/optical magnitudes, although a diminution in the NIR flux by $\sim 30$\% is observed.  In this section we will discuss what we can learn about the physical processes that power the X-ray and optical emission of the ULX on the basis of these observations.



\subsection{The big picture}

NGC 1313 X-2 is remarkable in that it is one of the half-dozen well-established PULXs, albeit the one with the faintest pulsations \citep{r6}.  The presence of a large bubble nebula around this object strongly argues for the occurrence of outflows and/or jets, an expectation of supercritical accretion models. In this work, we explored whether the two distinct modes of X-ray spectral behaviour it shows (as described in section~\ref{subsection:Xrayonly}) are also consistent with such models, derived originally from \cite{r10} and adapted for the presence of a magnetic neutron star by \cite{r102}.  We note that to fully understand this, exploration of the luminosity-temperature plane (e.g. \citealt{r1h}) via a continual monitoring in X-rays and at other wavelengths is necessary, rather than the two snapshots we have obtained; however we will proceed with this caveat in mind.

We notice that the change in X-ray spectral shape between the two observations occurs primarily through an increase in the 1.0-6.0 keV flux, while the lowest ($< 1$ keV) and highest energy ($> 6$ keV) components remain stable. Such an occurrence is qualitatively similar to at least two other ULXs (Holmberg IX X-1 and NGC 1313 X-1; \citealt{r22}; \citealt{r66}), and is difficult to reconcile with an intrinsic change in the mass accretion rate alone (see Table~\ref{Tab:Fits3}). This is supported by the seemingly invariant UV/optical and soft X-ray fluxes, which are (in part) emitted by the outermost regions of the disc, implying that the mass accretion rate through the outer disc remains steady over the three month timescale. As noted by \citet{r22}, the observed X-ray spectral transitions may instead be caused by a bulk precession of the thick inner disc and wind, plausibly due to the Lense-Thirring (or frame dragging) effect (\citealt{r23}). This naturally changes the observer's viewing angle to different parts of the accretion flow; a line-of-sight corresponding to the inner-most regions presents a view of the highest energy photons, which are geometrically collimated by the ultra-fast wind. Conversely, a view of outer wind photosphere reduces the hard X-ray flux, as the highest energy photons are Compton down-scattered by the outflowing material and/or scattered out of the line-of-sight. This picture is supported both by observational studies of how the soft X-ray spectral residuals correlate with luminosity (\citealt{r37}), in addition to numerical simulations of super-critical discs (\citealt*{r67}; \citealt*{r68}; \citealt*{r123}). 

Clearly, the precession scenario also predicts a stronger modulation in the hard X-ray band over long timescales, since the soft X-rays are less geometrically beamed and distributed over a larger solid angle, and should thus be less variable as the wind moves in/out of the observer's sight line. Indeed, recent studies have reported quasi-periodic hard X-ray variability in several (P)ULXs on weekly-monthly timescales (e.g. \citealt{r21}; \citealt{r69}; \citealt{r70}), invoking the above precession model as a possible explanation. In the particular case of X-2, \citet{r21} conclude that the variability is strongly periodic, although they do not consider red noise into their significance estimation, and the periodicity is not fully obvious in their folded light curve. In any case, \cite*{r71} discuss that a radiatively warped disc may undergo unstable precession, which need not induce strictly periodic modulations on the X-ray flux (although it is currently unclear whether this mechanism is preferred over the general relativistic Lense-Thirring effect for ULXs). 


The key question is why, if the X-ray emission changes, the optical does not?  Our analysis suggests that this may be due to the optical emission being dominated by the companion star, with a weaker contribution from an irradiated outer disc (although other factors may also play a role, as we discuss below). Indeed, this is perhaps not unexpected if the hard X-rays are collimated away from the accretion disc and (presumably) the companion star, and so cannot be reprocessed.  Hence, precession cannot affect the optical appearance of the PULX in the same way that it can affect the X-ray emission.

\subsection{The nature of the optical emission}

Our analysis of the optical light from the ULX counterpart shows it to be largely invariant between the two observational epochs, despite a change in the 0.3-10\,keV X-ray flux.  Using the magnitudes obtained in each of the different {\it HST\/} filters we can obtain a coarse spectrum of the counterpart, finding that its shape is broadly consistent with a B0III giant (if the NIR data point is ignored), but is also well explained by a power-law model (if the far UV data point is ignored), which is reminiscent of reprocessed X-rays.  We therefore attempted to fit the broadband X-ray - UV/optical SED with a disc reprocessing model, but this provided an unreasonably large reprocessed fraction ($> 80\%$ of the X-ray flux), although this is at least partly attributable to the unsuitability of the model to fitting super-Eddington accretion flows.  While a `slim' disc model is more justified in this respect, it does not consider X-ray reprocessing, and implicitly assumes that the outer disc is largely shielded from hard X-rays by the inflated inner disc. However, usage of this model implied that either unreasonably high accretion rates are required to account for the observed levels of UV/optical emission, or that an additional stellar component (dominating the UV/optical flux) is necessary. We emphasise that, while the light from the star dominating the counterpart appears probable from our results, we cannot fully rule out reprocessing since it is not accounted for by the slim disc model. 

Even if the optical light were to arise in reprocessed X-ray emission, changes in the reprocessed fraction may be difficult to observe in our case, given that the broadband X-ray flux increases only by $\sim 25\%$ between the observations probed by {\it{XMM-Newton}}. In the simple disc geometry applicable to sub-Eddington low mass X-ray binaries with a geometrically thin disc, $f_{\text{out}}$ can be predicted analytically. It is a function of the X-ray flux irradiating the outer disc $f_{\text{out}} \propto \big ( L_{\text{X}} \big )^{1/7}$ (\citealt{r59}), for a given compact object mass and disc size. This relation implies that a 25\% increase in $L_{\text{X}}$ should lead only to a 3\% increase in $f_{\text{out}}$, which cannot be demonstrated given the current quality of our data (see Table~\ref{Tab:Fits2}).  While this is consistent with the lack of variability we see, our accretion disc should not be in a geometrically thin configuration (as emphasised before).  Indeed, it is noted above that in supercritical discs one might expect the hard X-rays to be funnelled away from the disc, making large reprocessing fractions unlikely.


A final point to consider on reprocessing is that the longer wavelength emission may respond to changes in the X-ray flux with a time delay that is much larger than the duration of the snapshot {\it{HST}} observations ($15-20$ mins). This would clearly mask any correlated variability between these bands, although a crude estimate of the light-travel time between the X-ray and optically emitting regions of the disc may suggest that it is unlikely (see section~\ref{subsection:irradiation}). Indeed, our estimate does not account for the emergence of an ultrafast wind at accretion rates above Eddington (\citealt{r15}, \citealt{r18}). The lag between the X-ray and UV/optical flux could then be influenced by the photon diffusion timescales across the outflow, which will partly depend on the density of material within it.  Some evidence for such lags in ULXs is already emerging within the X-ray regime \citep{r115}.

Our modelling does however favour the optical emission arising in the companion star.  It is worth noting that the best-fitting stellar model reported here may not reflect the true donor spectral type, partly due to reasons outlined in \citet*{r41}. They demonstrate that the luminosity-temperature evolution of an isolated star can substantially differ from one in a binary. For example, an isolated star on the main sequence will tend to become more luminous as its radius expands. By contrast, a binary main sequence star (of the same initial mass) cannot expand above its Roche Lobe radius, and tends to decrease both in luminosity and temperature at the onset of Roche Lobe overflow. It will thus appear cooler and fainter than an isolated star at the same stage of its evolution, posing problems for a direct comparison of the observed magnitudes and colours with standard stellar templates. In addition, the donor star can also be irradiated by X-ray photons from the accretion flow, further affecting its brightness and colour (e.g. \citealt{r26}, although see comments above).  Hence, our attempts at identifying the stellar type may not be very accurate. 


Finally, we note that our work presents an intriguing link to other PULXs.  The first ULX for which the counterpart was unambiguously demonstrated to be stellar was NGC 7793 P13, also a PULX \citep{r26}.  A red supergiant companion has now been revealed for a second PULX, NGC 300 X-1 \citep{r116}, although attempts to uncover the counterparts for two more PULXs have proven inconclusive due to their distance, being heavily affected by obscuration and/or residing in crowded locations \citep{r117}.  Thus at least half of the PULXs have plausible stellar-dominated counterparts.  This contrasts strongly with the remainder of the ULX population where reprocessed X-ray emission remains a strong explanation for the optical counterpart \citep{r58}.  This could perhaps support the hard X-ray beaming in NS ULXs predicted by \cite{r118}, and seems consistent with the notion that NS ULXs would have smaller accretion discs than in their black hole equivalents \citep{r117}.

\subsection{What is the origin of the NIR variability?}

In almost all of our modelling, the NIR data point from the {\it HST\/} F125W filter appears to sit above the expected model emission, demonstrating a likely NIR excess in this ULX.  It is also notable that this filter shows the strongest evidence for variability in any of the {\it HST\/} data between the two observational epochs, with a $\sim 35\%$ decrease in NIR flux observed.  This is in stark contrast to an increase in X-ray flux between the same two epochs.

The fact that this variability occurs only in the NIR (whilst being absent in the UV/optical) immediately rules out an origin due to X-ray reprocessing, or arbitrary changes in the line-of-sight interstellar extinction.  It is also unlikely to come from the counterpart being a red supergiant star, which appears to be excluded by the observed SED\footnote{We note that \cite{r122} catalogue a possible red supergiant companion to X-2.  However, this object (their object B) is only just within their formal error region.  They also detect the object we regard as the counterpart (their object A) much closer to the centre of their error region, and derive magnitudes from the {\it HST\/} data consistent with ours.}.  We speculate that it may instead be attributed to transient jet emission, although we require additional evidence to confirm this. Classical Galactic black hole binaries (GBHBs) sometimes show NIR flaring episodes during accretion state transitions, from the low-hard state (featuring a persistent radio jet) to the high-soft/thermal dominant state (when jet emission is suppressed). These episodes are hypothesised to be caused by ejections of material from the hot inner flow.  Similar potentially explosive jet formation has also been seen in more than one ULX, for example in Ho II X-1 \citep{r119} and in a transient ULX in M31 \citep{r120}.

\section{Conclusions}



We detect clear X-ray spectral variations in NGC 1313 X-2 between {\it XMM-Newton\/} observations taken in December 2015, when 0.3-10\,keV {\it Swift\/} monitoring suggested the PULX was in a low flux, low variability mode, and subsequent observations in March 2016 when the PULX was in a higher flux and variability regime.  However, simultaneous optical/UV observations taken with {\it HST\/} demonstrate no correlated multi-wavelength variability, with the exception of an anti-correlation to the NIR emission of the counterpart: this fades as the X-ray emission rises between the two observation epochs.

We have discussed the X-ray and optical characteristics of this PULX in the context of a supercritical disc model, where a massive radiatively-driven wind is launched from the inner regions of the accretion disc as the disc becomes locally super-Eddington.  We find that the behaviour is consistent with this notion, with the changing X-ray properties plausibly the result of precession of the inner regions of the accretion flow.  The lack of optical variability is consistent with both a lack of reprocessing in the outer accretion disc/companion star due to the hard X-rays being collimated away from these media, and the optical counterpart being likely dominated by the stellar light; however this cannot explain previous reports of short-term ($\sim$ hours) optical variability in this object.  We also caution that our conclusions are drawn from a pair of snapshots; a much clearer picture of behaviour would emerge from longer-term simultaneous multi-wavelength monitoring of objects like X-2.

We also find evidence for other interesting phenomenology in this PULX, most notable of which is the variable NIR excess.  Possible explanations for this include a jet or a circumbinary disc, with the former perhaps the more plausible for this object.  

This work shows both the power and the pitfalls of simultaneous, multi-wavelength observations of interesting objects such as ULXs.  The simultaneity of the data puts clear constraints on the physics of the observed system; however to access such data requires the investment of observing time from the world's leading observatories, which is a scarce commodity, leading to limited datasets.  Obtaining the further monitoring data required to constrain the key behavioural characteristics of interesting objects such as the PULX NGC 1313 X-2 is therefore an expensive pursuit.  However, the continual development of new and future observatories with the capability for monitoring at a variety of wavelengths promises that this science will become more accessible with time, and so our understanding of objects such as PULXs should grow.

\section{Data Availability}
The data underlying this article will be shared on reasonable request to the primary author

\section*{Acknowledgements}
RS gratefully acknowledges the receipt of a studentship grant from the STFC ST/N50404X/1. TPR was funded as part of the STFC consolidated grant ST/K000861/1. RS is extremely grateful for the guidance provided by Andrew Dolphin on HST data analysis. This work is based on observations obtained with {\it{XMM-Newton}}, an ESA science mission with instruments and contributions directly funded by ESA Member States and NASA. The work is also based on observations made with the NASA/ESA Hubble Space Telescope, obtained at the Space Telescope Science Institute, which is operated by the Association of Universities for Research in Astronomy, Inc., under NASA contract NAS 5-26555. Our observations are associated with program \#14057. The authors would like to thank the {\it{XMM-Newton}}, {\it{HST}}, and {\it{Swift}} staff members for their help in planning and conducting these observations. EA acknowledges funding from the Italian Space Agency, contract ASI/INAF n. I/004/11/4. LZ acknowledges financial support from the Italian Space Agency (ASI) and National Institute for Astrophysics (INAF) under agreements ASI-INAF I/037/12/0 and ASI-INAF n.2017-14-H.0 and from INAF "Sostegno alla ricerca scientifica main streams dell'INAF" Presidential Decree 43/2018. RS acknowledges grant number 12073029 from the National Science Foundation of China. This work made use of data supplied by the UK {\it{Swift}} Science Data Centre at the University of Leicester. Lastly, we thank the anonymous referee for their useful comments and suggestions that helped improve the quality of the paper.

\bibliographystyle{mnras}
\bibliography{ref3_tpr} 
\bsp	

\label{lastpage}
\end{document}